\documentclass[
     reprint,
     superscriptaddress,
     amsmath,amssymb,amsfonts
     aps,
     pra,
     floatfix,
    ]{revtex4-1}

\usepackage{xfrac}
\usepackage{braket}
\usepackage{graphicx} \graphicspath{{PIC}}
\usepackage{verbatim}
\usepackage{etoolbox}
\usepackage[utf8]{inputenc}
\usepackage{mathtools}
\usepackage{soul,xcolor,colortbl}
\usepackage{hyperref} \hypersetup{colorlinks=true,citecolor=blue,linkcolor=blue,urlcolor=blue}
\usepackage{tabularx}
\usepackage{csquotes}
\usepackage{multirow}
\usepackage{longtable}
\usepackage{cellspace}
\usepackage{resizegather}
\usepackage{dcolumn}
\usepackage[normalem]{ulem}
\usepackage{siunitx}
\usepackage{physics}
\usepackage{pdfpages}
\usepackage{placeins}
\usepackage{natbib}
\usepackage{orcidlink}

\usepackage{bm}
\usepackage{array}
\newcolumntype{C}[1]{>{\centering\arraybackslash}p{#1}}\usepackage{soul}
\definecolor{Gray}{gray}{0.85}

\usepackage{color}
\usepackage{morefloats}

\definecolor{Gray}{gray}{0.9}
\definecolor{LightCyan}{rgb}{0.88,1,1}
\definecolor{green}{cmyk}{0,0,0,1}
\definecolor{blue1}{named}{black}
\definecolor{blue2}{named}{blue}

\renewcommand{\vec}[1]{\bm{\mathrm{#1}}}

\makeatletter
\AtBeginDocument{\let\LS@rot\@undefined}
\makeatother

\begin{document}

\title{{High-Throughput Prediction of Exfoliable Non-van der Waals Materials from a Universal Potential}}

\author{Tom Barnowsky\,\orcidlink{0000-0003-1626-4644}}
\affiliation{Theoretical Chemistry, Technische Universit\"at Dresden, 01062 Dresden, Germany}
\affiliation{Institute of Ion Beam Physics and Materials Research, Helmholtz-Zentrum Dresden-Rossendorf, 01328 Dresden, Germany}
\author{Carsten Timm\,\orcidlink{0000-0002-0279-0267}}
\affiliation{Institute of Theoretical Physics, Technische Universit\"at Dresden, 01062 Dresden, Germany}
\affiliation{W\"urzburg-Dresden Cluster of Excellence ctd.qmat, Technische Universit\"at Dresden, 01062 Dresden, Germany}
\author{Rico Friedrich\,\orcidlink{0000-0002-4066-3840}}
\email[Corresponding author, e-mail: ]{rico.friedrich@tu-dresden.de}
\affiliation{Theoretical Chemistry, Technische Universit\"at Dresden, 01062 Dresden, Germany}
\affiliation{Institute of Ion Beam Physics and Materials Research, Helmholtz-Zentrum Dresden-Rossendorf, 01328 Dresden, Germany}

\date{\today}

\begin{abstract}
\noindent

{
Exfoliation and cleavage create two-dimensional (2D) materials and surfaces with physical and chemical properties distinct from their bulk parents.
The rising class of non-van der Waals (non-vdW) 2D materials derived from non-layered crystals provides a fascinating platform, expanding the landscape of low-dimensional materials.
Current computational models, however, provide limited guidance: existing descriptors are largely tailored to vdW layered systems.
Here, we introduce a general framework predicting crystal cleavage and exfoliable 2D subunits directly from bulk structures.
At its core is a universal eXfoliation and Cleavage Potential (XCP) enabling large-scale screening of diverse materials at negligible computational cost.
Applying this approach, we obtain 44,030 cleavable surfaces and candidate non-vdW 2D materials from which we investigate -- according to our criteria -- 2,531 likely exfoliable ones using high-throughput density functional theory.
A large fraction of these candidates is found to be dynamically and thermodynamically stable, while showing negligible overlap with existing 2D materials databases.
Our study thus opens a systematic route to explore and design 2D materials with high chemical and structural diversity.
}

\end{abstract}

\maketitle

\section*{Introduction} \label{intro}

\noindent
Surfaces created by cleavage, exfoliation, or growth processes often determine the properties and functionalities of materials, for instance in the cases of topological insulators, heterogeneous catalysis, and electrochemistry.
In particular two-dimensional (2D) materials are essentially ``all surface" systems with nearly all atoms exposed due to their atomic-scale thickness.
Since the fabrication of single-layer graphene in 2004~\cite{Novoselov_Science_2004}, the landscape of 2D compounds has attracted great attention due to their unique physical and chemical properties.
They confine electrons within a plane, leading to pronounced surface effects and emergent quantum phenomena~\cite{Novoselov_Science_2007,Lei_nnano_2016,Deng_NNano_2016,Manzeli_NatRevMat_2017,Caldwell_nrm_2019,Gibertini_NNano_2019}.

Traditionally, such 2D materials are exfoliated from layered van der Waals (vdW) crystals, where weak interlayer interactions allow for mechanical separation of single layers.
However, this limitation has recently been shattered by the successful experimental exfoliation of non-vdW 2D materials derived from bulk crystals lacking an intrinsic layered structure.
In such systems, exfoliation involves the breaking of covalent or ionic bonds rather than vdW interactions, often producing surfaces terminated by unsaturated \enquote{dangling} bonds.
Prominent examples include MXenes, obtained by etching MAX phases~\cite{Anasori_NatRevMat_2017}, but also directly exfoliated systems such as 2D WO$_3$~\cite{Guan_AdvMat_2017}, hematene (2D Fe$_2$O$_3$)~\cite{Puthirath_Balan_NNANO_2018}, ilmenene (2D FeTiO$_3$)~\cite{Puthirath_Balan_CoM_2018}, and beyond~\cite{Jiang_NatSyn_2023,Kaur_AdvMat_2022}.
These systems feature strong chemical bonding, high surface reactivity, and often intrinsic magnetism~\cite{Puthirath_Balan_MatTod_2022,Gonzalez_2DM_2019}, making them promising for catalysis, adsorbate-driven tuning~\cite{Barnowsky_NanoLett_2024}, and functional superlattices and heterostructures~\cite{Zhao_Nature_2025,Nihei_ActaMat_2026}.

Initially, the selection of bulk materials for experimental exfoliation has been largely based on trial and error~\cite{Thakur_Science_2024}.
For traditional layered materials, highly successful data-driven approaches have been developed both to directly predict 2D materials \cite{Lyngby_npjcm_2022} and to guide experimental exfoliation from known bulk compounds~\cite{Ashton_PRL_2017,Mounet_AiiDA2D_NNano_2018} outlining thousands of candidates.
However, these approaches rely on special structural features such as
a simple binary overlap model of atomic spheres to identify vdW gaps in layered materials~\cite{Mounet_AiiDA2D_NNano_2018}.
This method classifies two atoms as bonded if their interatomic distance is below a certain threshold, and not bonded otherwise.
Such a geometric approach cannot capture exfoliability of non-layered non-vdW 2D materials where all atoms in the parent compounds are bonded in a three-dimensional network.

\begin{figure*}[ht!]
    \centering
    \includegraphics[width=.8\linewidth]{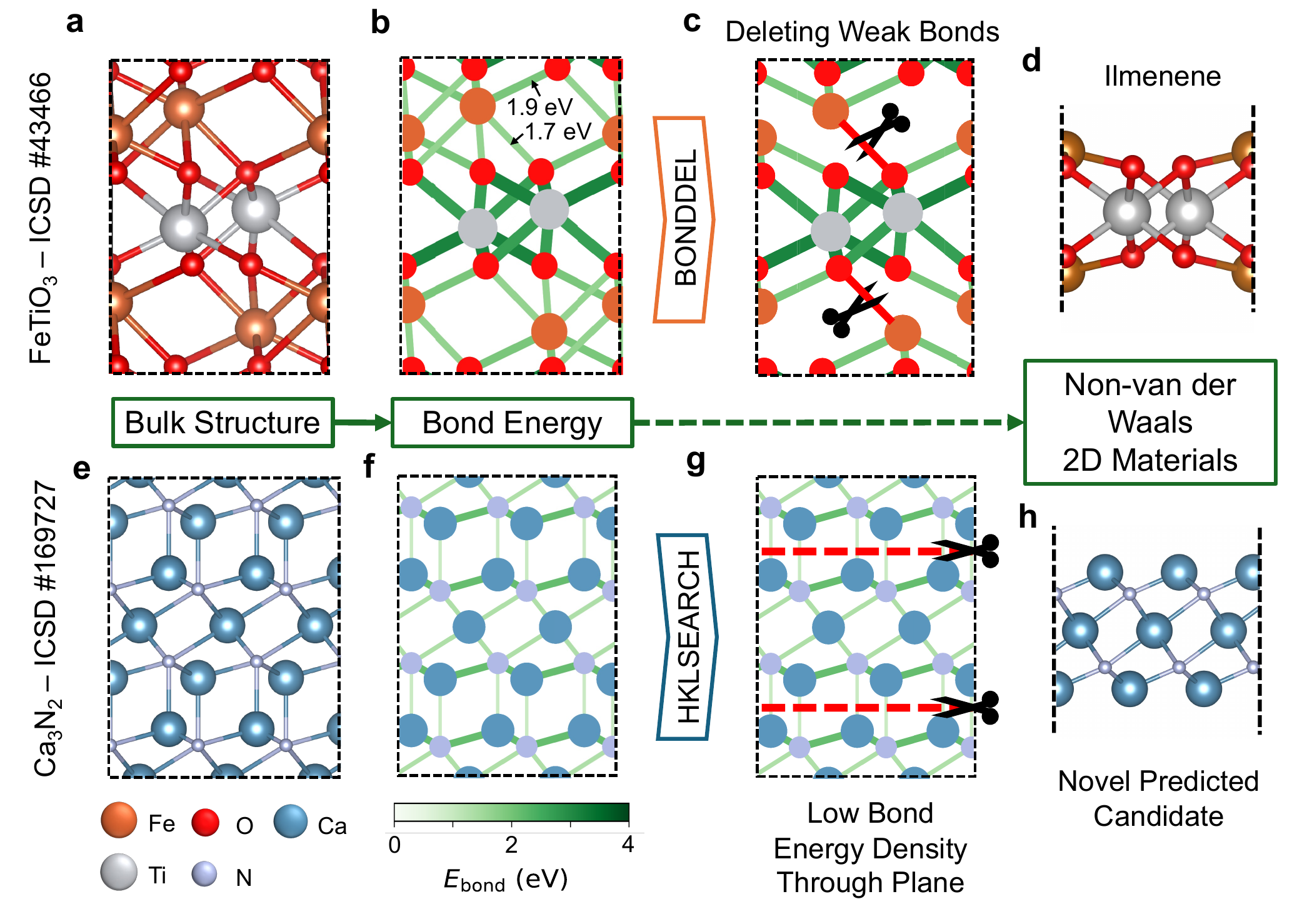}
    \caption{\small\textbf{Identifying 2D Subunits in Bulk Materials from Bond Energies.}
        Schematic illustration of how 2D sheets can be identified from bulk structures.
        As an example for the BONDDEL approach, the crystal structure of FeTiO$_3$ (ICSD \#43466) (\textbf{a}) is used to compute bond strengths using the {eXfoliation and Cleavage Potential} (XCP) model (\textbf{b}).
        The different Fe-O bonds are labelled by their respective strength.
        The weakest bonds in the structure are then cut until 2D subunits remain (\textbf{c}), leading to the previously experimentally realized~\cite{Puthirath_Balan_CoM_2018} {non-van der Waals} (non-vdW) 2D material ilmenene (\textbf{d}).
        The HKLSEARCH algorithm is demonstrated for Ca$_3$N$_2$ (ICSD \#169727) (\textbf{e}).
        Using the bond strengths provided by the XCP (\textbf{f}), the planes with lowest energy density (\emph{i.e.}, lowest surface energy) (\textbf{g}) are used to identify the new non-vdW 2D material (001) 2D Ca$_2$N$_3$ (\textbf{h}).
    }
    \label{fig1}
\end{figure*}

Thus, for this rising class of materials, an intuitive descriptor quantifying bond strength is essential.
While interatomic force constants have been used as a proxy to estimate interaction strength to find a range of new candidates~\cite{Bagheri_JPCL_2023,Bagheri_CM_2026}, cleavage and exfoliation are ultimately governed by thermodynamics and are thus energetic in nature.
{
Other approaches are limited to computationally expensive methods such as the calculation of elastic constants~\cite{Jia_npjcm_2021} or explicit electronic structure evaluations for each surface.
These are often preceded by simple pre-screening schemes based on packing ratios~\cite{Hu_NatComm_2023} or bond density~\cite{Zhong_npjcm_2025}, which are used to identify potentially relevant bulk facets.
In parallel, there have been conceptual approaches aimed at distinguishing non-vdW from conventional vdW materials~\cite{Ono_PRB_2025}, as well as studies focused on specific materials classes such as hematene-like structural prototypes~\cite{Friedrich_NanoLett_2022,Barnowsky_AdvElMats_2023} and MXenes~\cite{Bjork_Science_2024}.
}

Here, we present a universal description for cleavage and exfoliability of 2D sheets from arbitrary bulk crystals at minimal computational cost.
We parametrize a suitable potential to determine binding energy strength between individual atoms providing an enhanced understanding of the energy landscape and length dependence of bonds in the bulk phase.
Based on two different approaches either identifying weakly bonded crystallographic planes or from iteratively eliminating weak bonds, 2D subunits are determined.
Application of the scheme to the AFLOW-ICSD database~\cite{Esters_CMS_2023}, {a computational materials repository based on experimentally observed bulk crystal structures}, leads to the identification of 2,531 exfoliable non-vdW 2D candidates with a wide variety in chemical composition, structure, and properties.
{High-throughput calculations for the whole set confirm preferable small exfoliation energies for the large majority of candidates
and reveal that the main fraction is both dynamically and thermodynamically stable.
The identified 2D materials class shows negligible overlap with existing databases and thus spans a largely unexplored chemical space.
It includes interesting cases with hexagonal, square, and rectangular lattice symmetries exhibiting for instance metallic behavior or indirect-gap to direct-gap semiconductor transitions upon exfoliation, as demonstrated for several selected examples.}

\section*{Results} \label{results}

\textbf{Potential Model.}
To invoke exfoliation and cleavage, the interactions between atomic planes must be overcome.
Therefore, a meaningful yet efficient exfoliability descriptor should account for the strength of interatomic interactions within the bulk.

In the typical model of atoms connected by chemical bonds, the natural quantity to characterize binding strength is the bond energy $E_\mathrm{bond}$.
For efficiently determining the interaction strengths, a universal two-body potential is thus required, \emph{i.e.}, a potential parametrized for all species combinations.

To demonstrate the capabilities of this two-body potential ansatz, we formulate and fit a simple universal eXfoliation and Cleavage Potential (XCP):
\begin{equation}\label{eq:potential}
    V^{\text{Morse + Yukawa}}_{D,\alpha,r_0,C,\gamma}(r) = D \qty(1-e^{\alpha (r_0 - r)})^2 + \frac{C e^{-\gamma r}}{r},
\end{equation}
combining Morse and Yukawa terms.
For the Morse term -- typically used to describe covalent or metallic interactions -- the parameter $D$ corresponds to the potential well depth, $\alpha$ controls the width, and $r_0$ is the equilibrium distance.
The Yukawa potential captures ionic bonding, with the parameter $\gamma$ controlling the screening length.
The specific weights of both potential terms for each compound are determined via the parameters $C$ and $D$.
We note that the usage of analytical building blocks in the potential greatly enhances the data efficiency during fitting in contrast to determining the whole potential shape solely from large-scale general parametrization.

The five potential parameters ($D,\alpha,r_0,C,\gamma$) of our model are fitted separately for each individual species pair based on density functional theory (DFT) force and potential energy data from the AFLOW-ICSD dataset covering large parts of the periodic table~\cite{Esters_CMS_2023}.
{Further details as well as a verification of the developed interatomic potential and a discussing of the influence of different chemical environments are included in the Methods section.
We note that the AFLOW-ICSD reference values used for fitting are obtained from PBE($+U$) DFT calculations~\cite{PBE,Anisimov_Mott_insulators_PRB1991,LDAU1,LDAU2}.
Consequently, the resulting XCP inherits the systematic characteristics of this level of theory.
In particular, PBE is known to underestimate interactions with a significant dispersive contribution, implying that the XCP will likewise tend to underestimate bonding energies in such cases.
Due to the consistent training and application of the model, these effects are, however, expected to effectively cancel when predicting 2D candidates.
}

\begin{figure*}[ht]
    \centering
    \includegraphics[width=.9\linewidth]{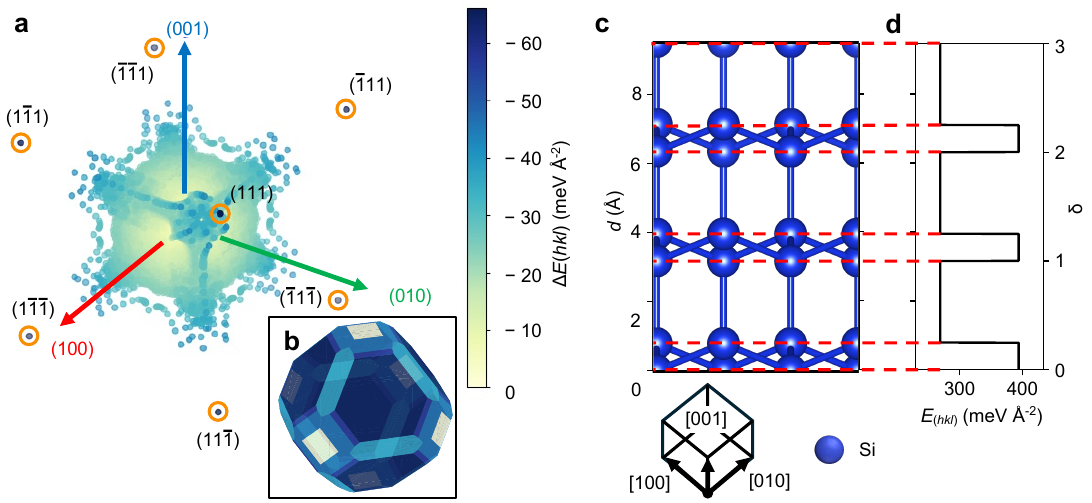}
    \caption{\small\textbf{Surface Energies, Exfoliation, and Cleavage Planes of Silicon.}
        (\textbf{a}) Three-dimensional plot showing the {energetic} favorability {$\Delta E (hkl)$} of Si planes in comparison to the maximum energy \{100\} planes.
        The quanitiy is represented by the radius and color (light = not favored, dark = highly favored).
        As a guide to the eye, the reciprocal lattice directions (100), (010), and (001), are indicated as arrows.
        The lowest energy \{111\} planes relevant for exfoliation and cleavage are marked by orange circles.
        (\textbf{b}) Wulff shape computed from these surface energies with the same orientation as in panel (\textbf{a}).
        (\textbf{c}) Crystal structure of Si aligned with the $(111)$ plane normal (= $[111]$ direct lattice direction) pointing upwards.
        (\textbf{d}) Energy profile of slicing the crystal at different positions.
        Panels \textbf{c} and \textbf{d} are aligned so that distances and atom positions match along the $(111)$ plane normal direction.
        The quantity $d$ measures the distance along this direction in \si{\angstrom} while $\delta$ is in units of the spacing of equivalent $(hkl)$ planes.
    }
    \label{fig2}
\end{figure*}

Universal two-body models based on physical constants rather than reference data, such as UFF~\cite{Rappe_1992_JCAS_UFF} or the original Lennard-Jones model based on equilibrium lengths and bond dissociation energies~\cite{LJ_OpenKIM}, were also tested but found to be insufficiently accurate for determining exfoliability and cleavage.
Higher-order fitted potentials including $N$-body contributions with $N>2$, such as ReaxFF~\cite{vanDuin_ReaxFF_jpca_2001} or MLIPs~\cite{Jacobs_cossms_2025}, cannot be simply reduced to purely two-body form.

Our XCP model enables the efficient estimation of bond strengths in arbitrary bulk structures, based on only atomic species and their interatomic distances, \emph{i.e.}, $E_\mathrm{bond} = V(r_{\mathrm{bond}})$.
In Fig.~\ref{fig1}, the bonding strengths determined using XCP are shown for ilmenite (FeTiO$_3$; ICSD \#43466) and calcium nitride (Ca$_3$N$_2$; ICSD \#169727).

\textbf{Exfoliability and Cleavage Identification.}
To identify cleavage planes and 2D sheets from the XCP bond energy description of three-dimensional (3D) structures, two complementary approaches are introduced:
(i) BONDDEL and (ii) HKLSEARCH schematically outlined in Fig.~\ref{fig1}\textbf{a}-\textbf{d} and \textbf{e}-\textbf{h}, respectively.

From a microscopic perspective, during the cleavage and exfoliation processes -- especially in liquid phase exfoliation (LPE) -- forces are applied to individual atoms.
In this case, the weakest bonds are broken first.
This concept is realized in the BONDDEL approach.
It proceeds by iteratively eliminating -- \emph{i.e.}, \enquote{deleting} -- the weakest bonds in the bulk structure until 2D substructures remain.
The working principle is visually evident from the bond strengths shown in Fig.~\ref{fig1}\textbf{b}, which leads to the experimentally known ilmenene sheets in Fig.~\ref{fig1}\textbf{c} when the weaker Fe-O bonds are broken as previously indicated by only considering varying bond lengths \cite{Friedrich_NanoLett_2022}.
The BONDDEL approach, however, does not yield 2D subunits for all bulk structures; for example, in a material where all bonds have equal strength such as Si, breaking them leads to disassembly into individual atoms rather than 2D sheets.
Hence, the cleavage planes of Si cannot be found using this approach.
However, non-planar cleavage{, \emph{i.e.}, when the cleavage surface is corrugated on an atomic scale rather than being a single planar cut,} observed in materials such as tungsten trioxide ($\alpha$-WO$_3$; ICSD \#27961)~\cite{Guan_AdvMat_2017}, requires BONDDEL to be identified correctly.

{
A bonding-network topology approach for identifying cleavable surfaces by analyzing the breaking of a selected number of bonds has been introduced previously~\cite{Paul_NPJCM_2021}.
This method identifies potential cleavage planes from structural information alone, by examining the scaling of bonding networks with supercell size.
The added value of our approach for cleavage lies in the additional quantification of bond strengths via the interatomic potential, which naturally provides a ranking of bonds and thereby indicates which bonds are most likely to be broken upon exfoliation.
As a result, the method enables a direct estimation of relative surface formation energies without requiring explicit DFT calculations for each cleavage configuration.
}

From a more macroscopic viewpoint, during the exfoliation process, forces are applied to the crystal, which makes the energetically weakest plane the most likely to be exfoliated.
The complementary HKLSEARCH algorithm thus analyzes the bond energy density across Miller index lattice planes, denoted as $E_{(hkl)}$.
This results in a static approximation of the surface energy that can be computed efficiently for arbitrary planes.
Miller indices with the lowest surface energy are deemed most likely to be cleavable and hence {potentially} exfoliable.

The case of bulk silicon, experimentally exfoliable along its \{111\} planes~\cite{Wang_mat_2019,Wang_JPCL_2020} -- the easy cleavage plane for the diamond structure -- is a prime example for the usefulness of this approach as outlined in Fig.~\ref{fig2}.
The favorability of the $\qty{111}$ planes is clearly demonstrated in Fig.~\ref{fig2}\textbf{a} via a three-dimensional visualization of the energetic preference $\Delta E(hkl)$  of all planes in comparison to the $\qty{100}$ planes with the highest surface energy.
They are evidently singled out in comparison to all other planes.

Another way to visualize the surface energies is the Wulff shape -- the equilibrium crystal shape, \emph{i.e.}, the shape of for instance nanocrystals of the material without external perturbation.
It is readily obtained from the XCP as shown in Fig.~\ref{fig2}\textbf{b} and corresponds well to the experimental reference data~\cite{Eaglesham_PRL_1993}.

The reason for the preferred cleavage planes lies in the low bond density through the $\qty{111}$ planes when the crystal is cut at the right position as highlighted in Fig.~\ref{fig2}\textbf{c} and \textbf{d}.
As for most real materials, the surface termination is crucial -- that is, different positions of the plane along its normal direction, here indicated by the parameter $\delta$, can lead to different bonding environments.
In Si, the (111) plane exhibits a strong variation of the surface energy depending on the location of the cut.
Only when the crystal is cleaved in a region of low bond-energy density (0.3 $\leq\delta\leq$ 1), low surface energy is realized.
Thus, to correctly capture the cut along the plane for any material, a minimization of the energy with respect to $\delta$ is crucial.

A final key aspect for quantifying exfoliability is that the sheets must not only be easily separated from the bulk but also remain structurally intact upon isolation.
To capture this, the exfoliability ratio $R$ between the energy required to exfoliate the sheet ($E_{\mathrm{3D \to 2D}}$) and the energy required to disassemble the 2D sheet into smaller subunits ($E_{\mathrm{2D \to <2D}}$) is introduced as a ranking factor:
\begin{equation} \label{ratio}
    R = \frac{E_{\mathrm{2D \to <2D}}}{E_{\mathrm{3D \to 2D}}}.
\end{equation}
Hence a high ratio ensures that the plane is stable in itself -- \emph{i.e.}, it does not easily break apart ($E_{\mathrm{2D \to <2D}}$ is large) -- while being easy to cleave from the bulk ($E_{\mathrm{3D \to 2D}}$ is low).
By using a ratio, the absolute bond strengths -- from weak vdW bonds to stronger covalent or ionic ones -- are irrelevant, only the inherent structural preference of planes and sheets determines their cleavability/exfoliability.
Thus, full generality over all materials classes is maintained.
This is conceptually similar to the ratio descriptor employing in-plane \emph{vs}. out-of-plane Young's moduli experimentally introduced for LPE~\cite{Backes_ACSNano_2019}.
The precise definition of the two energy values depends on the approach employed:
For HKLSEARCH, the surface energy of the selected plane and that of the lowest-energy perpendicular plane are used, whereas for BONDDEL, the minimum bond energies that must be overcome to obtain 2D and sub-2D fragments are considered.
An extended explanation of the two approaches, including implementation details, can be found in the Methods section.

\textbf{Validation from Experimental Data.}
To validate our XCP method, it is applied to a range of bulk materials of experimentally realized non-vdW 2D slabs~\cite{Guan_AdvMat_2017,Tai_advmat_2017,Li_acsnano_2018,Puthirath_Balan_NNANO_2018,Puthirath_Balan_CoM_2018,Yadav_AdvMatInt_2018,Puthirath_Balan_acsanm_2018,Wang_JPCL_2020,Zhang_angchem_2019,Xie_ACSAMI_2019,Liu_JMCA_2019,Feng_matter_2020,Puthirath_Small_2020,Guo_Nanoscale_2021,Ouyang_NML_2021,Hu_SmallStruct_2021,Goktuna_PSCE_2021,Peng_NChem_2021,Toksumakov_NPJ2DM_2022,Jiang_NatSyn_2023,Mahapatra_2DM_2023,Chattopadhyay_NL_2024,Su_advsci_2024} to predict the exfoliation planes.
Supplementary Table~1 includes the results from both approaches.
Bulk structures that are not in the AFLOW database are omitted from this statistics.
In total, a success rate of $\sim$\SI{80}{\percent} (23 out of 29) for identification is achieved.
This is highly remarkable given the complex and dynamic nature of the experimental exfoliation process mostly carried out in liquid-phase employing ultrasonication.
Indicative of the challenges is the fact that two experimental studies for pyrite (FeS$_2$; ICSD \#52372) report different exfoliation planes~\cite{Kaur_ACSNano_2020,Puthirath_JPCC_2021}.
Due to this ambiguity, this problematic case was also excluded from the benchmark.
The XCP method provides thus a highly efficient and accurate approach to predict candidates of non-vdW 2D materials.
We note that traditional vdW 2D materials from layered 3D counterparts are identified reliably by default, as their large vdW gaps lead to weak interactions and hence weak interlayer binding energies.

Considering that the choice of non-layered bulk systems to be exfoliated in experiments was previously mainly based on intuition rather than quantitative insight, the observed ratios $R$ are frequently close to unity.
In this regime, the accuracy of both identification approaches is expected to be lower than at higher ratios, as small inaccuracies in the calculated binding energies have a stronger effect on the predicted exfoliability.
Hence, the developed XCP method can be expected to exhibit even higher predictive power for system with high values of $R$.

\begin{figure*}[ht!]
    \centering
    \includegraphics[width=.95\linewidth]{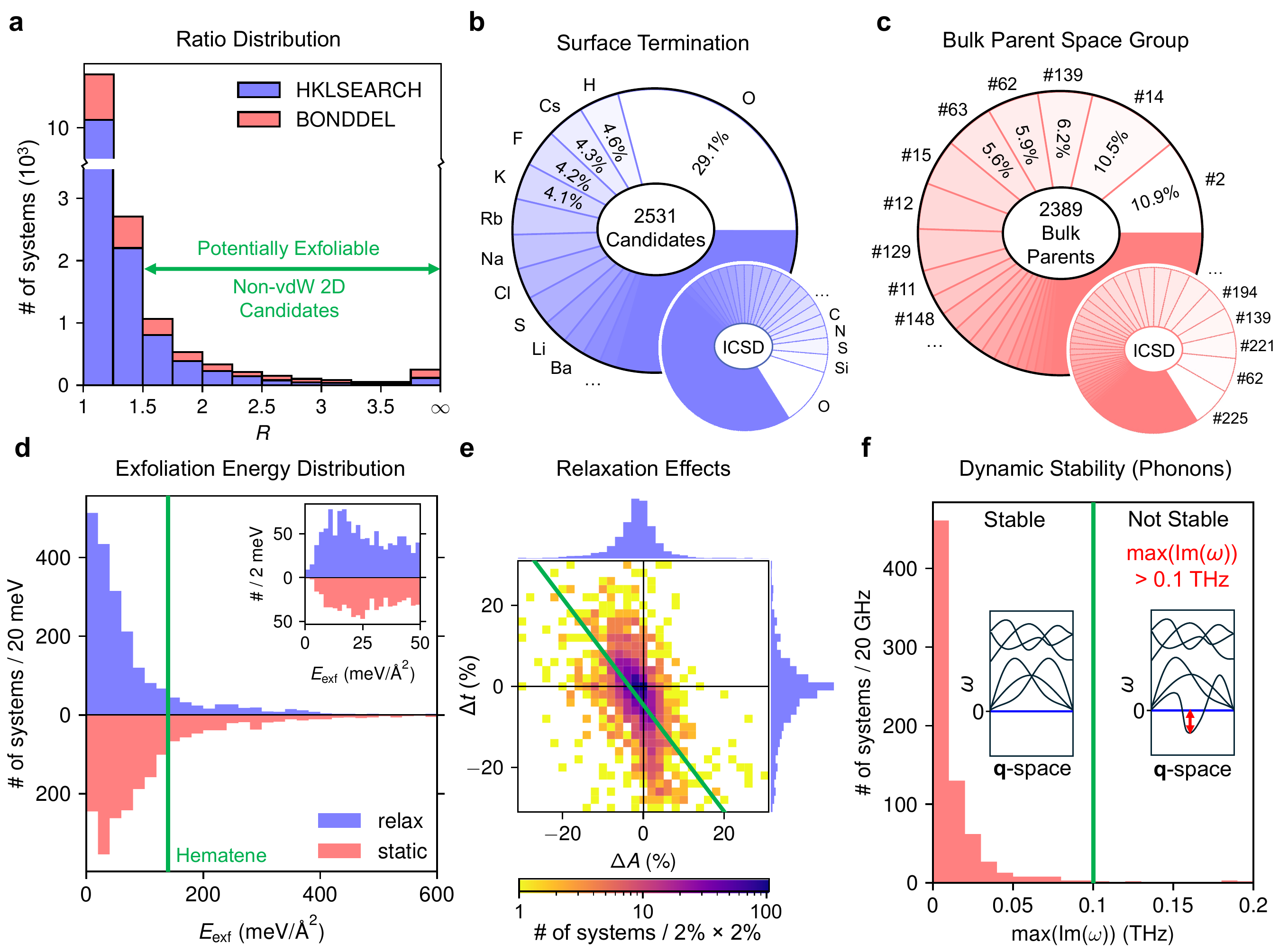}
    \caption{\small\textbf{Predicted 2D Candidates.}
        (\textbf{a}) Distribution of exfoliability ratios from the HKLSEARCH and BONDDEL approaches.
        Bars resulting from both methods are stacked.
        The rightmost bar includes all structures with ratios above $3.75$.
        The range of investigated potentially exfoliable candidates ($R > 1.5$) is outlined.
        (\textbf{b}) Pie chart of relative frequencies of different surface-terminations among the identified 2D candidates.
        The data for the species distribution among compounds
        in the AFLOW-ICSD is included as inset.
        (\textbf{c}) Pie chart showing the distribution of bulk parent space groups (by index number) giving rise to 2D sheets.
        The space group distribution in the AFLOW-ICSD is included as inset.
        (\textbf{d}) Distribution of exfoliation energies for the candidates.
        Relax/static values correspond to computations including/excluding structural relaxation of the obtained 2D sheets.
        $E_\mathrm{exf} = \SI{140}{\milli\eV\per\angstrom^2}$ computed for hematene, \emph{i.e.}, 2D (001) Fe$_2$O$_3$ \cite{Friedrich_NanoLett_2022}, is indicated by the green vertical line.
        The inset highlights the region around the maximum.
        {
        (\textbf{e}) 2D histogram illustrating the structural changes of the candidates upon exfoliation.
        The marginals show the distributions along the two dimensions: the relative thickness change $\Delta t$ and the relative area change $\Delta A$.
        The main trend, determined via linear regression, is indicated by the green line.
        (\textbf{f}) Distribution of the maximum imaginary phonon frequencies of the candidate materials.
        The visualization is restricted to the range from 0 to \SI{0.2}{\tera\Hz} to emphasize the stability threshold at \SI{0.1}{\tera\Hz} (highlighted in green).
        Insets show schematic phonon dispersions for a dynamically stable and an unstable material, with the maximum imaginary frequency marked in red.
        }
  }
    \label{fig3}
\end{figure*}

\textbf{Predicting Novel Candidates.}
{To identify new 2D materials, we focus on 3D bulk parent systems known from experiments.
Hence, b}oth approaches are applied to all unique bulk structures from the AFLOW-ICSD dataset excluding layered materials (or assemblies of one- or zero-dimensional subunits held together by vdW forces) according to the standard descriptor introduced by Mounet~\emph{et al.}~\cite{Mounet_AiiDA2D_NNano_2018}.
In total, the XCP method yields {44,030} cleavable surfaces/2D structures derived from {24,566} bulk 3D crystals.
Note that the number of identified 2D structures exceeds the number of bulk systems since a single bulk compound can give rise to multiple 2D sheets, for example if it consists of alternating monolayers or if the two identification approaches yield complementary results.
For further analysis, especially for computing exfoliation energies $E_\mathrm{exf}$~\cite{Jung_NanoLett_2018}, only systems whose composition is conserved upon exfoliation are considered (excluding, \emph{e.g.}, MXenes).
{The compositions of both the bulk and 2D systems are compared for each candidate individually.}

The distribution of these {17,897} {composition-conserving} systems as a function of the exfoliability ratio $R$ is shown in Fig.~\ref{fig3}\textbf{a}.
As expected, the majority of systems exhibits low ratios close to unity, indicating that they are very isotropic and not readily exfoliable.
However, a significant number of several thousand systems show higher ratios, suggesting that they are promising candidates for non-vdW 2D materials.
Consequently, {out of these,} we decided to perform a detailed high-throughput analysis {including structural relaxations} on the {2,531} {unique composition-conserving} systems with $R > 1.5$.

{
Since we classify each surface identified by the XCP approach as ``cleavable'' these composition-conserving high-ratio materials can be considered ``likely exfoliable'' non-vdW 2D candidates.
This does not imply that systems with smaller values of $R$ cannot be exfoliated into non-vdW sheets, as demonstrated by a range of experiments discussed above{, \emph{e.g.}, $R = 1.06$ for hematene and $R = 1.13$ for silicene}.
However, due to the finite accuracy of the XCP model to estimate relative bond strengths in different directions, the predictive power of the present model is expected to increase for larger values of $R$.
Therefore, the cutoff at 1.5 is chosen conservatively to ensure that the candidates identified here are highly likely to be practically exfoliable -- potentially even with less effort than many of the previously reported experimental examples.
}

Figure~\ref{fig3}\textbf{b} shows the distribution of surface terminations among the identified materials in comparison to the presence of these elements in AFLOW-ICSD parents.
While the most common elements in the database largely mirror their abundance in earth's crust (O, Si, ...), for the non-vdW 2D materials the trend proposed in Ref.~\cite{Friedrich_NanoLett_2022} is confirmed, with the most frequent terminating species in low oxidation states $\pm1$ or $\pm2$.
With {around \SI{30}{\percent}}, most slabs are terminated by oxygen, usually in the $-2$ state, which is also the most common element among compounds in the AFLOW-ICSD, albeit with a smaller overall fraction.
It is followed by {H$^{+}$ (\SI{4.6}{\percent}), Cs$^{+}$ (\SI{4.3}{\percent}), F$^{-}$ (\SI{4.2}{\percent}), and K$^{+}$ (\SI{4.1}{\percent})}.
{
However, this does not necessarily mean that experiments on non-vdW 2D materials will always be able to directly observe these terminations.
Typical exfoliation techniques, especially LPE, work in a chemically complex environment that may lead to surface passivation by other species or functional groups.
}

The space-group distribution of bulk parent materials giving rise to non-vdW 2D materials (Fig.~\ref{fig3}\textbf{c}) shows a shift to groups with a low number of symmetry operations, \emph{i.e.}, low-index space groups.
Leading is the anisotropic space group {\#2 ($P\overline{1}$) with \SI{10.9}{\percent} followed by \#14 ($P2_1/c$, \SI{10.5}{\percent}), \#139 ($I4/mmm$, \SI{6.2}{\percent}), \#62 ($Pnma$, \SI{5.9}{\percent}), and \#63 ($Cmcm$, \SI{5.6}{\percent})}.
The AFLOW-ICSD has a different focus with \#225 ($Fm\overline{3}m$) and \#62 ($Pnma$) being most common, \emph{i.e.}, more highly symmetric space groups.
Lower symmetry structures promote exfoliation due to inherently elevated anisotropy and coordination with broken symmetry leads to weaker bonds in certain directions enabling low energy cleavage planes.
We note that the 2D slabs themselves cannot be classified according to space groups~\cite{Fu_2DM_2024}.

The {2,531} candidates {belong} to {all} five possible 2D Bravais lattices.
Most of the systems have a rectangular lattice ({\SI{43.3}{\percent}}) followed by oblique ({\SI{18.7}{\percent}}) and square ({\SI{17.4}{\percent}}) variants.
Among the different lattices, there are {504} rectangular (REC), {295} oblique (OBL), {121} square (SQR), {112} hexagonal (HEX), and {73} rectangular-centered (RECC) oxidic sheets.
Notably, the  experimental fabrication of square lattice oxides has very recently been reported~\cite{Zhou_ResearchSquare_2025}.

We also note that the XCP method can identify MXenes from the parent MAX phases as verified in a few example cases.
It reliably assigns both the MX sheet to be extracted as well as the A-layer eliminated during etching.
A detailed investigation of such systems is, however, beyond the scope of this work.

{In addition, the XCP approach can also be valuable in the context of certain growth processes.
While kinetics plays a decisive role during growth, often the final {shapes of particles and platelets} are determined by the minimization of surface energy as outlined by the Wulff construction, which can thus be captured by XCP.}

\begin{figure*}[ht!]
    \centering
    \includegraphics[width=\linewidth]{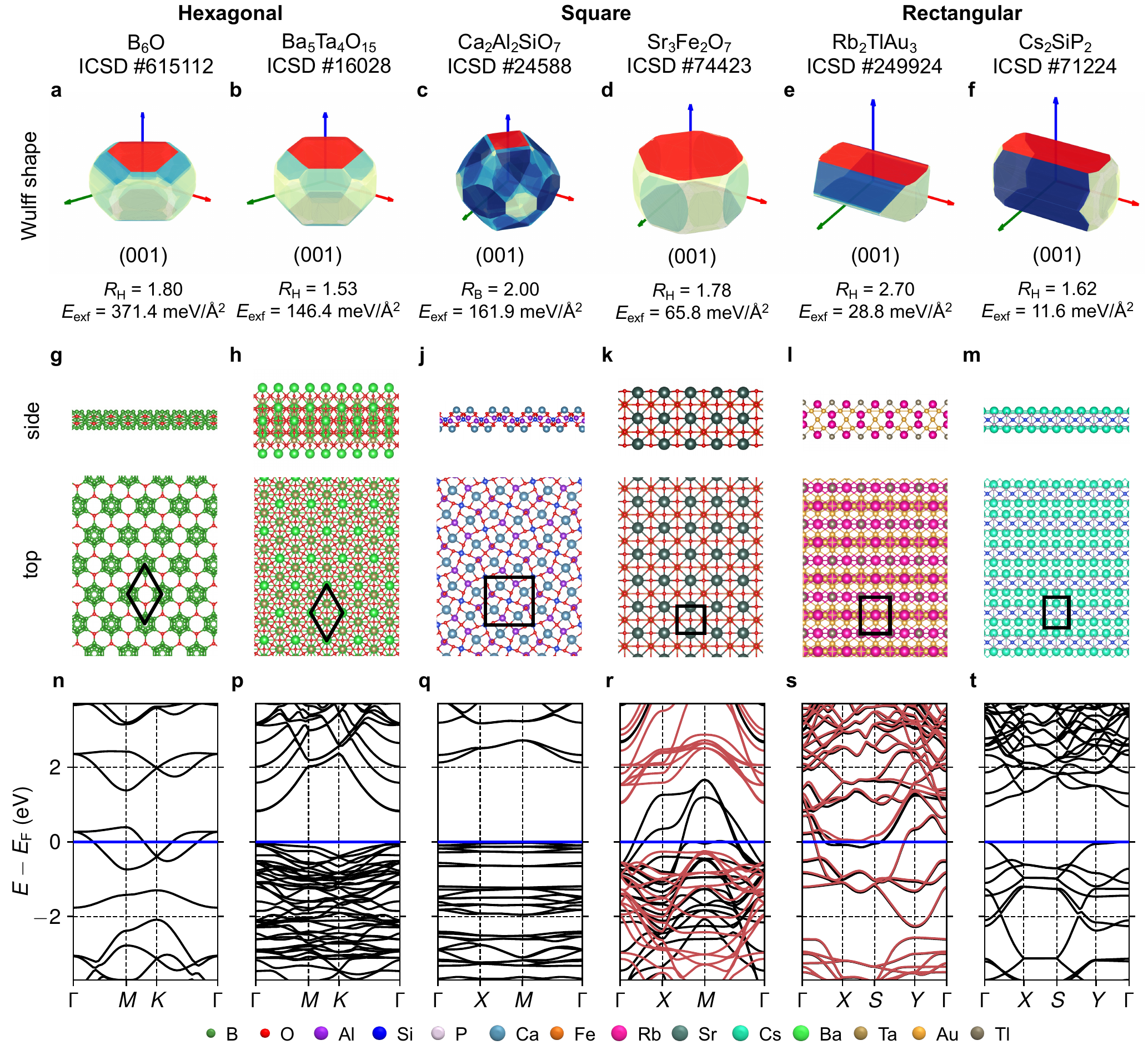}
    \caption{\small\textbf{Atomic and Electronic Structure of Selected 2D Candidates.}
      Example non-vdW 2D materials identified by the XCP method including the exfoliation plane, chemical formula, ICSD number, exfoliability ratio according to the HKLSEARCH (H) and BONDDEL (B) approaches, and calculated exfoliation energy.
    {
        (\textbf{a}-\textbf{f}) Wulff shapes of bulk structures yielding the non-vdW 2D candidate determined using XCP.
        Different facets are colored according to their surface energy (darker = lower) for better visibility.
        The exfoliation plane is rotated upwards and indicated in red.
        Reciprocal lattice directions of the bulk crystal are shown as red ($\vec b_1$), green ($\vec b_2$), and blue ($\vec b_3$) arrows.
        }
        (\textbf{g}-\textbf{m}) Top and side views of 2D structures with in-plane unit cells are indicated by the black frames.
        The systems are sorted vertically according to their lattice type, \emph{i.e.}, hexagonal, square, and rectangular.
        (\textbf{n}-\textbf{t}) Corresponding band structures { displayed on a joint energy scale}.
    }
    \label{fig4}
\end{figure*}

\textbf{Properties of Identified 2D Candidates.}
Fig.~\ref{fig3}\textbf{d} shows the distribution of computed {PBE($+U$)} DFT exfoliation energies obtained by comparison with the corresponding bulk AFLOW entries.
Values taking into account geometry optimization of the 2D slab are shown in the upper half while static \enquote{as sliced} $E_\mathrm{exf}$ are shown in the lower part.
The latter give an indication of the energy scale required to initiate the cleavage process~\cite{Barnowsky_AdvElMats_2023}.
The exfoliation energies range from close to zero to around \SI{0.5}{\eV\per\angstrom^2} in few cases.
The vast majority ({\SI{85.9}{\percent}} for relax) lies below \SI{140}{\milli\eV\per\angstrom^2}, the value for the prototypical non-vdW 2D material hematene~\cite{Puthirath_Balan_NNANO_2018,Friedrich_NanoLett_2022}, indicating that most candidates should be experimentally accessible.

{
Another common measure of thermodynamic stability is the convex hull distance energy, $\Delta E_\mathrm{hull}$, \emph{i.e.}, the energetic distance of the 2D system to the convex hull of stability -- a measure of energy per atom rather than per area~\cite{curtarolo:art144}.
We estimate this distance here with respect to the bulk parent structures taken from the AFLOW-ICSD dataset which have been realized experimentally.
$\Delta E_\mathrm{hull}$ can be related to the exfoliation energy $E_\mathrm{exf}$ via the atomic area density of the unrelaxed 2D sheet.
While $E_\mathrm{exf}$ is the more natural quantity for 2D materials derived from bulk via exfoliation, $\Delta E_\mathrm{hull}$ can be straightforwardly extended to stacks of $N$ ``monolayers'' by the estimate, $\Delta E_\mathrm{hull}(N) \approx \Delta E_\mathrm{hull}(1) / N$.
The resulting distributions of $\Delta E_\mathrm{hull}$ for different $N$ are shown in Supplementary Fig.~1.
A commonly used criterion for thermodynamic stability is $\Delta E_\mathrm{hull} < \SI{100}{\milli\eV\per atom}$~\cite{Gjerding_2DMats_2021}.
Based on this threshold, a fraction of \SI{49.0}{\percent} of the candidates is thermodynamically stable in the monolayer limit, which increases rapidly to \SI{70.0}{\percent} for bilayer stacks.
}

The distribution {of exfoliation energies and convex hull distances} visibly shifts to lower values when taking into account structural optimization, leading to large energy gains for these systems in contrast to traditional vdW 2D systems.
The relevance of this effect for the exfoliation of some non-vdW 2D materials was already highlighted in Refs.~\cite{Friedrich_NanoLett_2022,Barnowsky_AdvElMats_2023} and taking into account surface relaxations can lower $E_\mathrm{exf}$ by up to a factor of four.

{
The structural relaxations are substantial and follow a clear trend, as shown in Fig.~\ref{fig3}\textbf{e} for the full dataset.
The relative thickness change of the slab, $\Delta t$ (with thickness defined as the vertical distance between the outermost atoms), is anti-correlated with the in-plane area change $\Delta A$.
These structural changes are in particular strong compared to vdW materials, which remain largely unchanged when extracted from their bulk parent structures.
A set of examples demonstrating characteristic relaxation behavior of non-vdW 2D sheets is provided in Supplementary Fig.~2.
A typical mechanism is the flattening of the broken coordination of surface ions, as exemplified by 2D (010) MgSiO$_3$ (ICSD \#172736), which exhibits an area increase of \SI{2.74}{\angstrom^2} ($\Delta A = \SI{15}{\percent}$) combined with a thickness reduction of \SI{1.33}{\angstrom} ($\Delta t = \SI{-30}{\percent}$).
This can be attributed to the loss of out-of-plane bonding pulling surface ions towards the interior of the slab.
In contrast, (111) K$_2$O (ICSD \#641282) shows the opposite behavior, with an increase in thickness of \SI{0.44}{\angstrom} ($\Delta t = \SI{23}{\percent}$) and a decrease in in-plane area of \SI{3.24}{\angstrom^2} ($\Delta A = \SI{-18}{\percent}$).
Here, structural relaxation is driven by enhanced separation of K ions at opposite surfaces, leading to increased buckling.
For more structurally complex materials, the same general trends persist, although a reduction to simple coordination arguments becomes increasingly difficult.
For example, (001) SiO$_2$ (ICSD \#171739) undergoes a long-range reconstruction, resulting in a large area increase of \SI{53.20}{\angstrom^2} ($\Delta A = \SI{40}{\percent}$) while preserving similar local atomic environments.
Conversely, (001) OsP$_4$ (ICSD \#647710) exhibits a near-complete reorganization of the P atoms around Os upon exfoliation, accompanied by a thickness increase of \SI{0.87}{\angstrom} ($\Delta t = \SI{23}{\percent}$) and an area reduction of \SI{9.01}{\angstrom^2} ($\Delta A = \SI{-18}{\percent}$).
}

{
As it cannot be excluded that some of the 2D candidates predicted in this work have previously been found in the context of other high-throughput 2D materials studies, we investigate the overlap with other databases in detail.
Significant overlap with vdW-materials-based datasets such as the Materials Cloud 2D Database (MC2D)~\cite{Campi_ACSNano_2023} and 2DMatPedia~\cite{Zhou_sdata_2019} is unlikely, since layered compounds were excluded from the input structures to which the descriptor was applied.
There might be greater potential for overlap with the Computational 2D Materials Database (C2DB), which is largely based on bottom-up generation of 2D candidates~\cite{Gjerding_2DMats_2021}.
However, in practice we observe zero overlap with the MC2D, only three materials are found by the present work and 2DMatPedia as well as two duplicate materials with respect to the C2DB are identified.
Comparisons are performed using a stoichiometry and total energy based scheme outlined in the Methods section.
These results indicate that non-vdW 2D materials form a new class orthogonal to conventional 2D materials, thereby substantially expanding the accessible design space for 2D compounds.
}

{
A key criterion for predicting materials is their dynamical stability.
To evaluate this, the phonon dispersion $\omega(\vec q)$ for all DFT-relaxed structures are computed on a dense $\vec q$-point mesh using the mace-mh-1 foundation model MLIP~\cite{Batatia_ARXIV_2025}.
A material is considered dynamically stable if no imaginary frequencies appear anywhere in the first Brillouin zone.
In practice, we quantify this by evaluating the maximum imaginary component.
To account for numerical noise in the context of soft modes \cite{Radescu_PRM_2019}, a structure is classified as dynamically unstable only if this value exceeds \SI{0.1}{\tera\Hz}.
This criterion is illustrated in Fig.~\ref{fig3}\textbf{f}.
The distribution of maximum imaginary frequencies exhibits a pronounced peak near zero, with a finite spread extending up to \SI{0.1}{\tera\Hz}.
This peak corresponds to the dynamically stable majority, comprising \SI{64.4}{\percent} of the systems.
In contrast, unstable structures show significantly larger imaginary frequencies that are broadly distributed over several THz, resulting in a lower probability density.
As such, instabilities do not preferentially occur near zero frequency.
The observed fraction of stable systems lies between the near-\SI{100}{\percent} stability expected for vdW 2D materials, whose structures closely resemble their naturally layered bulk counterparts, and the \SI{28.4}{\percent} stability reported for C2DB composed primarily of generated structures~\cite{Gjerding_2DMats_2021}.
}

Figure~\ref{fig4} showcases a selection of predicted {dynamically stable} non-vdW 2D systems {starting from the Wulff shape of their bulk parents, over their relaxed slab geometry, to} their electronic structure -- underlining the variety of this rising class of materials.
{The Wulff shape is particularly interesting as for a flat appearance with dominant facets intuitively good cleavage properties can be expected.
The examples} are sorted vertically according to their lattice type, \emph{i.e.}, including hexagonal, square, and rectangular lattice systems.

{Remarkably, B$_6$O (ICSD \#615112) transitions from a direct semiconductor in its bulk phase to a 2D metal upon exfoliation of its (001) plane, with the states at the Fermi level originating from boron-derived orbitals associated with the two out-of-plane B--B bonds per B$_{12}$ icosahedron that are cut with respect to the bulk.
The bulk indirect-gap insulators Ba$_5$Ta$_4$O$_{15}$ (ICSD \#16028), Ca$_2$Al$_2$SiO$_7$ (ICSD \#24588), and Cs$_2$SiP$_2$ (ICSD \#71224)} exhibit an emergent direct band gap ($\Gamma \to \Gamma$) upon exfoliation.
This behavior highlights the potential {of these materials} for optoelectronic applications.
In contrast, {Sr$_3$Fe$_2$O$_7$ (ICSD \#74423), a half-metal in the bulk, retains its ferromagnetic and half-metallic} character in the square (001) monolayer.
The heavy-element compound Rb$_2$TlAu$_3$ (ICSD \#249924) likewise preserves the metallic nature of its bulk phase when reduced to the rectangular (001) sheet.

Sheets identified by BONDDEL are particularly intriguing, as they frequently exhibit non-planar surface terminations, such as the (001) Ca$_2$Al$_2$SiO$_7$ structure illustrated in Fig.~\ref{fig4}\textbf{j}.
{
In contrast to systems identified via HKLSEARCH, where the exfoliation plane and its symmetry-equivalent facets clearly dominate the Wulff shape as evident in Figs.~\ref{fig4}\textbf{a}--\textbf{b} and \textbf{d}--\textbf{e}, this feature is much less pronounced for systems with corrugated surfaces (Fig.~\ref{fig4}\textbf{c}).
The reason is that such cuts cannot be represented by planar $(hkl)$ facets within the XCP surface-energy formalism.
}

These examples reveal the elevated diversity and potential use of the predicted candidates for functional applications compared to prior non-vdW 2D systems that have so far been largely limited to hexagonal semiconductors similar to hematene.
As for many non-vdW 2D materials, the new candidates frequently contain magnetic transition-metal ions, offering potential for spintronic applications due to strong spin polarization associated with magnetic ions at or near the surface of the 2D sheet.

\section*{Discussion} \label{discussion}
\noindent
This work introduces a universal computational approach for predicting cleavage of crystals and for the identification of (non-vdW) 2D materials based on only crystal structure information at negligible computational cost.
At its heart is the eXfoliation and Cleavage Potential (XCP) model providing a highly efficient yet accurate parametrization of interatomic interactions, \emph{i.e.}, chemical bonds.
Two complementary approaches detect weakly bonded planes and iteratively isolate connected 2D subunits:
HKLSEARCH finds favored lattice planes, and BONDDEL identifies surfaces via microscopic bond breaking.
Compared to costly electronic-structure calculations, the XCP method enables rapid, database-scale screening of materials.

XCP predicts exfoliability and cleavage across all crystal structure types, reproducing experimentally realized non-vdW sheets and paving the way to a huge number of candidates.
It explains preferred slab terminations and links exfoliable (cleavage) planes in systems such as silicon to high in-plane/out-of-plane surface-energy ratios.
We have identified overall {44,030} cleavable surfaces and potential 2D systems, out of which {2,531} conserve the composition of the bulk parent material and are likely to be exfoliable -- increasing the pool of known non-vdW 2D systems by two orders of magnitude.
The set of candidates includes systems from all possible 2D Bravais lattices including rectangular, oblique, and square variants.
High-throughput density functional calculations confirm small exfoliation energies for the large majority of candidates and showcase a variety of appealing properties.
{
In addition, a large fraction of the systems is found to be both dynamically stable, based on phonon calculations, and thermodynamically stable according to convex hull distance criteria.
}
They reveal semiconducting sheets with a direct band gap in contrast to their parent bulk materials as well as first non-vdW 2D metals.
{
Comparison with existing 2D materials databases shows negligible overlap, highlighting that the present approach accesses a largely unexplored class of materials.
}

By construction, the method reaches beyond earlier approaches limited to purely vdW interlayer interactions.
It thus generalizes previously reported correlations between bond-length anisotropy and exfoliable planes to a next level bond-energy framework.
The highly intuitive XCP approach thereby not only uncovers a vast and diverse space of novel non-vdW 2D materials but also demonstrates that it is useful in a larger context in surface science: it is capable of predicting surface energies and extractable subunits of any material.

\section*{Methods}

\textbf{Neighbor Determination.}
When working with interatomic potentials in periodic systems, infinite neighbor sums must be truncated.
A basic approach is to consider only neighbors within a maximum radius $r_{\mathrm{cut}}$.

While this is very intuitive, it is not sufficient for determining the neighbors relevant for chemical bonding since typical bond lengths differ between different atomic environments.
Using a species- or environment- dependent cutoff is problematic as it will usually not be symmetric under exchange of bonding partners,
\begin{equation}
    j \in \text{Neigh}(i) \quad\nLeftrightarrow\quad i \in \text{Neigh}(j).
\end{equation}
This is visualized in Supplementary Fig.~3 for the example of an Fe--O--H bonding arrangement.

To resolve this, in our work, neighbors are determined by checking whether the nearest neighbor radii of two atoms overlap, \emph{i.e.}, if the distance between both atoms is smaller than the sum of their nearest neighbor radii,
\begin{equation}
    j \in \text{Neigh}(i) \quad\Leftrightarrow\quad r_{ij} < C\qty(r^{\mathrm{NN}}_i + r^{\mathrm{NN}}_j).
\end{equation}
This approach is inherently symmetric and can be tuned using a scaling factor $C$.
Different values of $C$ determine which order of neighbors are found, \emph{e.g.}, in the case visualized in Supplementary Fig.~3 $C = 0.8$ would find the Fe--O and O--H bond (first neighbors) while $C = 1.1$ finds also the second neighbor interaction Fe--H.
To capture the attractive first-neighbor interaction as well as in ionic crystals the potentially repulsive second neighbor one, we use $C = 1.1$.

\textbf{Potential Model.}
The XCP model described above was trained on the AFLOW-ICSD dataset including DFT calculations for ICSD structures.
Duplicates are filtered out by energy per atom: structures with the same composition and energy within \SI{1e-4}{\eV} are considered duplicates.
In addition, only species combinations realized for at least 20 entries are used in the training data.
This results in a total of 7,700 model parameters, \emph{i.e.}, 1,540 species combinations with five parameters each using a Morse+Yukawa XCP potential.

For each system $\zeta_i$ in the AFLOW-ICSD database, energy $L^E$ and force $L^F$ deviations (losses) are minimized,
\begin{align}
    L^E(i) &= \frac{1}{|\zeta_i|}|E^{\mathrm{XCP}}(\zeta_i) - E^{\mathrm{ref}}_i|,\\
    L^F(i) &= \frac{1}{|\zeta_i|}\sum_{\text{Atom } j \in \zeta_i} ||\vec F^{\mathrm{XCP}}_j(\zeta_i)||,
\end{align}
where $|\zeta_i|$ is the number of atoms in the unit cell.
XCP energies are calculated via Eq.~(1), while forces are analytically derived from the two-body potentials and compared to the zero reference force of the ground-state AFLOW-ICSD structures.

The total loss function is then defined as the root mean square (RMS) deviation of energy and force:
\begin{equation}\label{eq:loss}
    L = \operatorname{RMS}_{i}\ L^E(i) + \Omega\ \operatorname{RMS}_{i}\ L^F(i),
\end{equation}
with the hyperparameter $\Omega$ tested and fixed at $\SI{0.5}{\angstrom}$ for most stable convergence.
Optimization uses the L-BFGS-B quasi-Newton approach with an \SI{80}{\percent}/\SI{10}{\percent}/\SI{10}{\percent} train/validation/test split with a termination at the minimum of the validation loss as shown in Supplementary Fig.~4\textbf{a}.

The final potential reaches a test loss $L_\mathrm{test} = \SI{0.97}{\eV\per atom}$, with the energy contribution $L_\mathrm{test}^E = \SI{0.67}{\eV\per atom}$.
For octahedral coordination with six neighbors, this corresponds to a per-bond error of $\sim\SI{0.1}{\eV}$.
Energy deviations for all systems in the dataset are shown in Supplementary Fig.~4\textbf{b}.

{
The loss function and training set are not additionally balanced with respect to the number of occurrences of specific species combinations.
As shown in Supplementary Fig.~5, the distribution of bond types is biased toward common species, meaning that frequently occurring combinations contribute more strongly to the fit.
As a consequence, these bond types are represented with higher accuracy in the resulting potential.
Given that the species distribution in the AFLOW-ICSD dataset can be considered representative of real materials, this imbalance aligns the model with physically relevant configurations.
In comparison to a balanced approach, it therefore improves accuracy for previously unseen materials whose compositions follow the same underlying distribution as the training data, while less common species combinations are correspondingly less accurately described.
}

\textbf{Potential Validation.}
A sanity check to validate the XCP is provided by comparison of bonding radii.
For this purpose, the equilibrium radii $r^0$ of the potential for each species combination were determined,
\begin{equation}
    r^0_{ij} = \operatorname{argmin}_{r>0} V(r,Z_i,Z_j).
\end{equation}
As shown in Supplementary Fig.~6, these are in good agreement with the sum of the covalent radii of the given species combination~\cite{Cordero_DT_2008}.

{
Constructing the potential energy such that it solely depends on atomic species and pairwise interatomic distances allows the model to describe interactions across the full periodic table with a comparatively small number of parameters.
This is in contrast to approaches that explicitly incorporate oxidation states or other local environment descriptors.
This simplicity, however, also limits its accuracy, as bond strengths are influenced by the local chemical environment.
At the same time, many of these local effects are implicitly encoded in a quantity the potential depends on: the interatomic distance.
This is because the local environment affects not only bond strength but also the characteristic bond length.
Provided that the underlying structure is obtained from an accurate method, these geometric features already reflect the local chemical environment to large extent.
As a result, a potential that depends only on species and distance can still distinguish between different chemical environments.
This is exemplified in Supplementary Fig.~7 for the O--Fe interaction, where the common Fe oxidation states +2 and +3 separate in energy rather well based solely on their characteristic bond lengths~\cite{Friedrich_PRM_2021,Friedrich_CCE_2019}.
Consequently, the model is well suited for evaluating bond strengths in given structures, while it is not intended for structure optimization or molecular dynamics, where the lack of explicit environment dependence would tend to homogenize bond lengths across different chemical environments.
}

\textbf{Preferred Lattice Planes (HKLSEARCH).}\label{hklsearch}
The HKLSEARCH approach first evaluates $E_{(hkl)}$ for a large set of planes, \emph{i.e.}, all integer Miller indices with $|h|,|k|,|l| < 10$.
This is achieved by extending the broken-bond model originally developed by Mackenzie \emph{et al.}~\cite{Mackenzie_1_jopacos_1962,Mackenzie_2_jopacos_1962}, which accurately captures the number density of bonds intersecting $(hkl)$ surfaces with the potential energy of individual bonds as determined using XCP.
This enables the calculation of the minimal cleavage energy for any given Miller plane $(hkl)$,
\begin{equation}\label{eq:emin}
    \begin{aligned}
        E_{(hkl)} &= \min_{\delta \in[0,1)}\qty{\sum_m \ \sum_{i \in \mathrm{Neigh}(m)} V^{\mathrm{XCP}}(\vec b_i) \, n_{(\vec h, \delta)}(\vec d_m, \vec b_i)}\\
       &= \min_{\delta \in[0,1)} E_{(hkl)}(\delta),
    \end{aligned}
\end{equation}
where $\delta$ parametrizes the position of the cleavage plane between neighboring equivalent $(hkl)$ lattice planes, $\vec h$ is the plane normal, $\vec d_m$ is the position of the $m$-th basis atom, $\vec b_i$ is the $i$-th bond vector emanating from that atom, and $n_{(\vec h, \delta)}(\vec d_m, \vec b_i)$ is the bond density crossing the plane, as defined in Ref.~\cite{Mackenzie_2_jopacos_1962}.
The plane exhibiting the lowest cleavage energy is predicted to be the most likely to exfoliate.
The optimal cutting position $\delta$ is determined using explicit minimization with a \SI{}{10^{-3} \angstrom} step size.

In the range $\delta \in [0,1)$ multiple non-equivalent parallel 2D sheets may be contained.
They are all considered results of the approach with ratio $R$, even though each could present a different energy barrier for exfoliation.
However, this can only be finally evaluated using an electronic-structure method such as DFT, as on the XCP level they are bound equally strongly.

Energetic preferences of $(hkl)$ planes versus the highest surface energy plane of the material are computed as
\begin{equation}
    \Delta E(hkl) = \max_\qty(h'k'l') E_{(h'k'l')} - E_{(hkl)}.
\end{equation}

To quantify exfoliability within this approach, Eq.~(2) corresponds to the in-plane/out-of-plane surface energy ratio,
\begin{equation}\label{eq:hklratio}
    R = \frac{E_{\perp}}{E_{\parallel}} = \frac{\min_{(h'k'l') \perp (hkl)} E_{(h'k'l')}}{E_{(hkl)}}.
\end{equation}
A material is considered easy to exfoliate along a plane if it is simple to cleave parallel to the plane (low $E_{\parallel}$) but difficult to cleave perpendicular to it (high $E_{\perp}$), resulting in a large $R$.

The technical challenge that arises when evaluating $\min_{(h'k'l') \perp (hkl)} E_{(h'k'l')}$ is that calculations are only performed for a finite set of Miller indices.
Often, no plane in the set is exactly perpendicular to the reference $(hkl)$, \emph{i.e.}, the dot product of the plane normals is not precisely zero.
To address this, the scalar product tolerance is gradually increased until a sufficient number of points (here ten) are found to reliably determine the minimum perpendicular cleavage energy.

Wulff shapes are defined by
\begin{equation}
    \mathcal{W} = \left\{ \vec x \in \mathbb{R}^3 \,\middle|\, \vec x \cdot \vec h \leq E_{(\vec h)} \text{ for all } \vec h  \right\},
\end{equation}
where $\vec h$ are the $(hkl)$ lattice plane normals.

\textbf{Microscopic Bond Breaking (BONDDEL).}\label{bonddel}
After determining bond energies from the XCP, one can identify 2D materials by iteratively breaking/deleting the weakest bonds -- setting their bond energy to zero -- until 2D subunits remain.
This approach is referred to as BONDDEL.

Two steps are necessary for this approach.
In the first step, it determines which atoms remain bound to each other while accounting for periodic boundary conditions, and in the second step, it determines the dimensionality of each remaining connected subunit.

To identify connected atoms, the structure is represented as an undirected graph.
Nodes correspond to atoms and edges correspond to bonds, with edge weights given by the bond energies.
Periodic boundary conditions are naturally included by adding edges connecting to periodic images of atoms in neighboring cells.
Thus, in a crystal with a single-atom basis, the node representing this atom has edges only to itself connected to all periodic images.

Connected subunits are identified by traversing the graph using depth-first search (DFS).
Starting from an arbitrary atom, all atoms reachable via non-zero weight edges are assigned to the same subunit.
The process is repeated starting from atoms that have not yet been reached until all atoms are assigned to some subunit.
The dimensionality of each subunit is determined by constructing a $2 \times 2 \times 2$ supercell of the original unit cell restricted to atoms belonging to that subunit.
The number of connected replicas in the supercell then determines the dimensionality $D$ of the subunit.

Starting from the weakest bonds, the weights of increasingly strong bonds are iteratively set to zero.
While initially the structure is one single periodic 3D connected unit, this process will lead to the disassembly into smaller and potentially lower-dimensional units.

The necessary bond energy required to break the bulk into 2D subunits is denoted by $E_{D=2}$.
If progressively stronger bonds are broken, the 2D slab eventually disassembles into $D<2$ subunits -- 0D single atoms in the extreme case -- with the corresponding energy denoted by $E_{D<2}$.
From these two energies, the exfoliability ratio can be defined analogously to Eq.~(2),
\begin{equation}\label{eq:bondratio}
    R = \frac{E_{D<2}}{E_{D=2}}.
\end{equation}

Miller indices can still be assigned to 2D sheets identified by this approach, making use of the ability to easily identify planes using HKLSEARCH.
This is independent of the employed potential if one introduces sufficient vacuum, \emph{i.e.}, by placing only a single sheet in a $2 \times 2 \times 2$ supercell.

{
\textbf{2D Material Comparisons.}
Uniqueness within the set of 2,531 exfoliable candidates, as well as overlap with existing 2D materials databases, is assessed using a hashing scheme that assigns each material a unique string of the form \texttt{<stoichiometry>\_xcp\_<XCP total energy>}.
Here the energy is expressed as a value in \si{\eV\per atom} up to two significant figures.
While this does not constitute a structural comparison at the level of more advanced tools such as AFLOW-XtalFinder~\cite{curtarolo:art170}, the combination of stoichiometry and total energy makes accidental collisions between distinct 2D materials highly unlikely.
This approach is applied to both unrelaxed and relaxed structures: the former to avoid redundant DFT calculations on equivalent candidates, and the latter to enable direct comparison with ground-state structures reported in external materials databases.
}

\textbf{Exfoliation Energy Calculations.}
The exfoliation energies $E_{\mathrm{exf}}$ are calculated based on the work of Jung \emph{et al.}~\cite{Jung_NanoLett_2018},
\begin{equation}
    E_{\mathrm{exf}} = \frac{1}{A_{\mathrm{slab}}}\qty(E_{\mathrm{slab}}-\frac{N_{\mathrm{slab}}}{N_{\mathrm{bulk}}}E_{\mathrm{bulk}}),
\end{equation}
where $E_{\mathrm{slab/bulk}}$ are the energies of the slab and bulk structures and $N_{\mathrm{slab/bulk}}$ is the respective number of atoms.
The normalization uses the in-plane surface area $A_{\mathrm{slab}}$ of the unoptimized 2D unit cell.

All DFT calculations are performed using the Vienna \emph{Ab-Initio} Simulation Package (VASP)~\cite{vasp_prb1993,Bloechl1994a,vasp_JPCM_1994,vasp_prb1996,vasp_cms1996} with the PBE($+U$) functional~\cite{PBE,Anisimov_Mott_insulators_PRB1991,LDAU1,LDAU2}, employing the AFLOW software~\cite{Divilov_HighEntAlloyMat_2025} and its standard settings~\cite{Calderon_cms_2015} {(for details see Section~VII in the Supplementary Informtion)} to ensure compatibility with the AFLOW database.
{
Reciprocal space integration grids for relaxations were determined using 2D-KPPRA = 400 while the subsequent static calculation, used to evaluate the total energy and DOS, employs 2D-KPPRA = 2,500.
The band-structure calculations shown in Fig.~\ref{fig4} were performed based on the static geometry and charge density.
The on-site $U$ values listed in Supplementary Tables 2 and 3 are employed during geometry optimization as well as static and band structure calculations.
}

{
Phonon calculations were performed using the mace-mh-1 MLIP~\cite{Batatia_ARXIV_2025}.
This most recent MACE model was chosen as it was developed using cross-learning strategies and is therefore expected to perform well for structures that deviate from the typical PBE($+U$) bulk training sets, such as the 2D sheets considered here.
A second MLIP based on a similar philosophy, SevenNet-omni~\cite{Park_JCTC_2024}, was also tested and yields consistent trends.
To ensure accurate forces, all DFT optimized structures were re-relaxed with the MLIP prior to the phonon calculations.
This introduces a mismatch $\epsilon$ between the DFT-relaxed geometry and the MLIP ground state, which we quantify using the structure comparison implemented in AFLOW-XtalFinder~\cite{curtarolo:art170}.
Structures with $\epsilon < 0.1$ are considered equivalent; consequently, if the MLIP-relaxed structure satisfies this criterion, the MLIP is deemed sufficiently accurate for the corresponding candidate.
This is the case for a total of \SI{65.3}{\percent} of candidates that are then carried over to the statistics shown in Fig.~\ref{fig3}\textbf{f}.
A benchmark assessing the reliability of MLIP-predicted phonon dispersions for a representative set of non-vdW 2D candidates, compared with DFT \emph{ab initio} results from Ref.~\cite{Barnowsky_AdvElMats_2023}, is provided in Supplementary Fig.~8.
There a clear trend emerges: in four of the six cases, the MLIP-relaxed structure exhibits $\epsilon < 0.1$ to the corresponding DFT-relaxed structure, consistent with the fraction observed across the full dataset.
For these four materials, the MLIP phonon dispersions are in good agreement with the DFT results -- particularly in the low-frequency region that is most relevant for identifying dynamical instabilities.
Phonon dispersions were computed within the finite-displacement framework as implemented in the atomic simulation environment (ASE)~\cite{Larsen_JPhysCM_2017}.
In-plane supercells were constructed dynamically to ensure a minimum separation of \SI{40}{\angstrom} between periodic images.
The sampling of $\omega(\vec q)$ was performed on a dense in-plane grid (2D-KPPRA = 25,000) to reliably capture the maximum imaginary frequency.
}

For the systems shown in Fig.~\ref{fig4}, the preferred magnetic order was determined by explicitly evaluating the energies of all possible magnetic configurations within the primitive unit cell, or within a $2 \times 2$ in-plane supercell for systems with only one magnetic center per primitive cell.

\section*{Data Availability}
The associated data is made publicly available via the Rossendorf Data Repository (RODARE) at \verb|https://doi.org/10.14278/rodare.4175|~\cite{Barnowsky_Rodare_2025}.

{
\section*{Code Availability}
The methods developed in this work can be employed interactively for any input crystal structure at} \verb|https://xcp.hzdr.de|.
In addition, the software implementing the approaches is openly available on RODARE at \verb|https://doi.org/10.14278/rodare.4180|~\cite{Barnowsky_FINDSLAB_Rodare_2025}.

\newcommand{\Ozolins}{Ozoli{\c{n}}{\v{s}}}

\section*{Acknowledgments}
The authors thank Steve Schmerler, Anastasiia Nihei, Chen-Chen Er, Moritz Leucke, and Kornelius Nielsch for fruitful discussions
{This work was supported by generous grants of CPU time from} the HZDR Computing Center, HLRS Stuttgart, the Paderborn Center for Parallel Computing PC$^2$, and TU Dresden ZIH.
{We thank David Pape and the HZDR Computational Science department for their support in creating and maintaining the} \verb|https://xcp.hzdr.de| {website}.

\section*{Funding}
R.~F. acknowledges funding by the German Research Foundation (DFG), project FR 4545/2-1, ID 513655797, collaborative research center SFB 1415, project C08, project ID 417590517, and for the \enquote{Autonomous Materials Thermodynamics} (AutoMaT) project by Technische Universität Dresden and Helmholtz-Zentrum Dresden-Rossendorf within the DRESDEN-concept alliance.
C.~T. gratefully acknowledges financial support by DFG through collaborative research center SFB 1143, project A04, project ID 247310070, and through W\"urzburg-Dresden Cluster of Excellence ctd.qmat, EXC 2147, project ID 390858490.

\section*{Author contributions} \label{contribs}

R.~F. initiated and supervised the project.
T.~B., C.~T., and R.~F. conceptualized the potential based exfoliability and cleavage method.
T.~B. performed the potential parametrization, the software implementations, data-analysis and post-processing as well as all calculations.
T.~B. wrote the initial draft of the manuscript, {which was subsequently revised and refined by T.~B. and R.~F}.
All authors contributed to the writing and discussion of the final manuscript.

{
\section*{Competing Interests}
The authors declare no competing interests.
}

\clearpage
\FloatBarrier
\onecolumngrid
\begin{center}
  \Large
  \textbf{
  Supplementary Information
  }
\end{center}

\setcounter{figure}{0}
\setcounter{table}{0}

\section{Supplementary Methods}

\subsection{The AFLOW Standard}
The parameters employed in the Vienna Ab-initio Simulation Package (VASP)~\cite{vasp_prb1993,Bloechl1994a,vasp_JPCM_1994,vasp_prb1996,vasp_cms1996} were chosen according to the AFLOW standard~\cite{Calderon_cms_2015}. 
This includes the selection of pseudopotentials (PPs) as listed in Supplementary Table~\ref{tabS2}, as well as on-site interaction parameters applied to selected $d$- and $f$-shell elements. 
Specifically, effective Hubbard $U$ parameters are used for $d$-elements, while both $U$ and $J$ parameters are employed for $f$-elements~\cite{Anisimov_Mott_insulators_PRB1991,LDAU1,LDAU2}, as detailed in Supplementary Tables~\ref{tabS3} and \ref{tabS4}. 

Each PP provides a recommended minimum plane-wave cutoff energy (ENMAX). 
To ensure reliable convergence, the cutoff energy in each calculation is set to 1.4 times the maximum ENMAX among the PPs used.

AFLOW uses an automatic approach to generate $\vec k$-point grids for bulk material, ``$\vec k$-points per reciprocal atom'' (KPPRA).  
For the in-plane reciprocal space integration grids ($N_1 \times N_2 \times 1$) used for 2D cells a modified version was used.
The parameter 2D-KPPRA is defined such that $\text{2D-KPPRA} \leq N_\mathrm{atoms} N_1 N_2$, where $N_\mathrm{atoms}$ is the number of atoms in the unit cell. 
The distribution of sampling points $N_i$ along the two in-plane reciprocal lattice vectors $\vec b_i$ is chosen to best match the reciprocal lattice geometry, \emph{i.e.}, to approximately satisfy $||\vec b_1|| / N_1 = ||\vec b_2|| / N_2$.

\section{Supplementary Figures}

\vspace{-.4cm}
\begin{figure*}[ht]
    \centering
    \includegraphics[width=\linewidth]{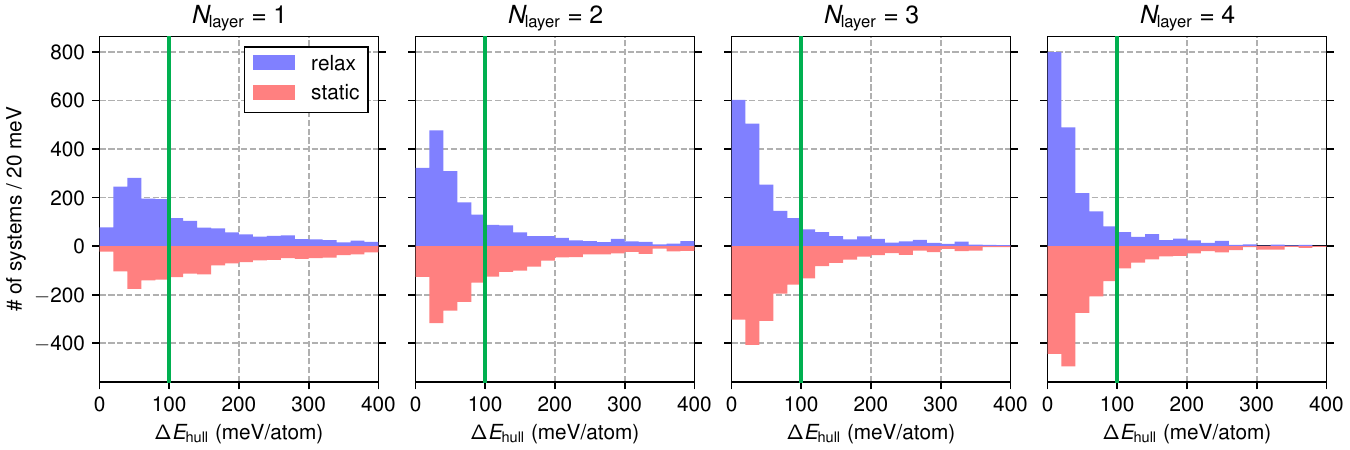}
    \caption{\small\textbf{Layer dependence of the convex hull distance energy.}
    	Distribution of convex hull distance energies $\Delta E_\mathrm{hull}$ for the candidates as a function of the number of layers $N_\mathrm{layers}$. 
    	``Relax'' and ``static'' refer to calculations including and excluding structural relaxation of the derived 2D sheets, respectively. 
    	The common stability criterion $\Delta E_\mathrm{hull} = \SI{100}{\milli\eV\per atom}$ is indicated by the green vertical line.
	}
    \label{figS1}
\end{figure*}

\begin{figure*}[ht]
    \centering
    \includegraphics[width=.9\linewidth]{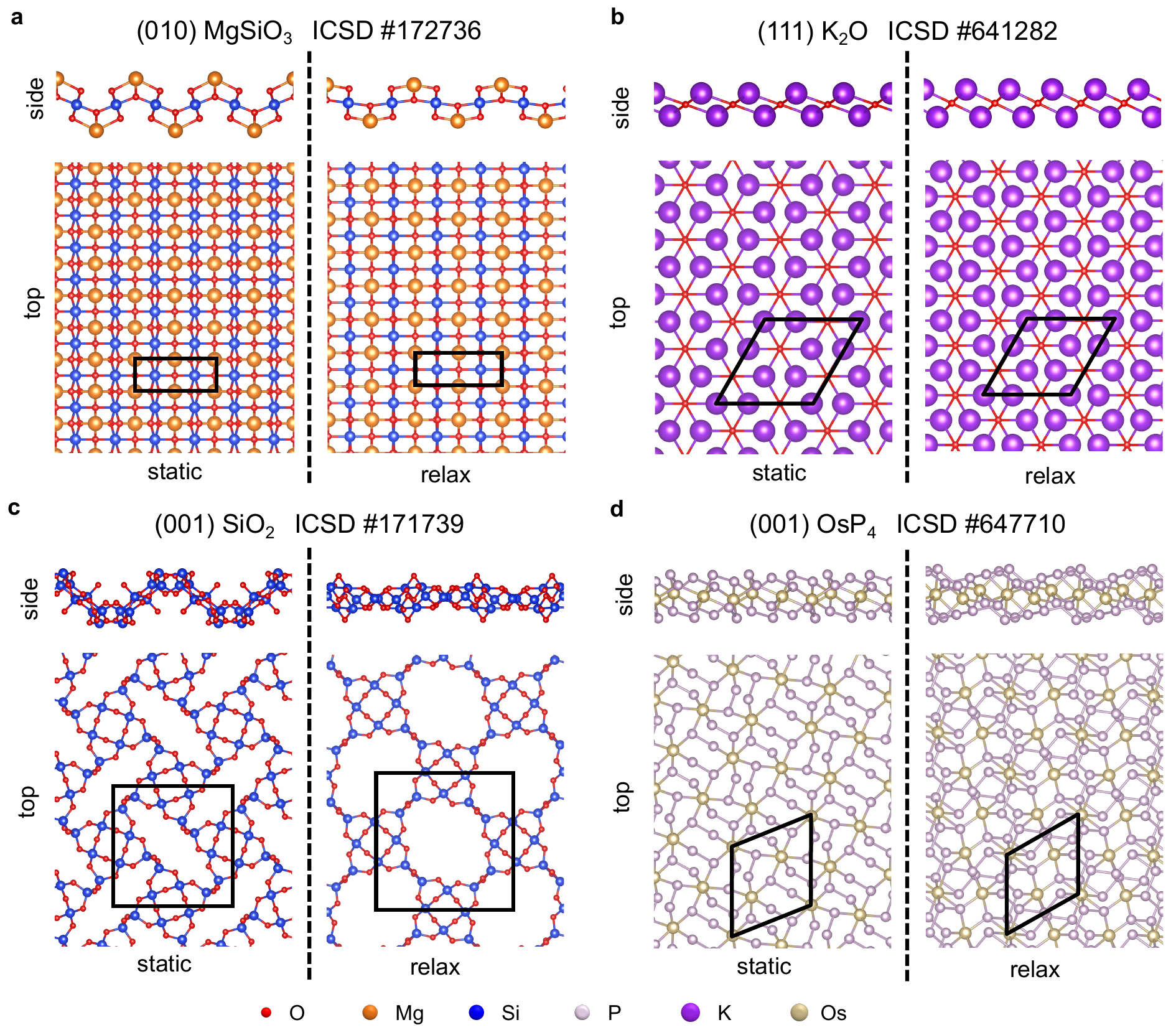}
    \caption{\small\textbf{Examples for Relaxation Effects.}
    Top and side views comparing 2D structures immediately after cleavage (``static'') and after structural optimization (``relax''). 
	For each material, all views are displayed on a common scale to highlight relaxation-induced changes in area and thickness. 
	Black frames indicate the in-plane unit cells.
    }
    \label{figS2}
\end{figure*}

\begin{figure*}[ht]
    \centering
    \includegraphics[width=.36\linewidth]{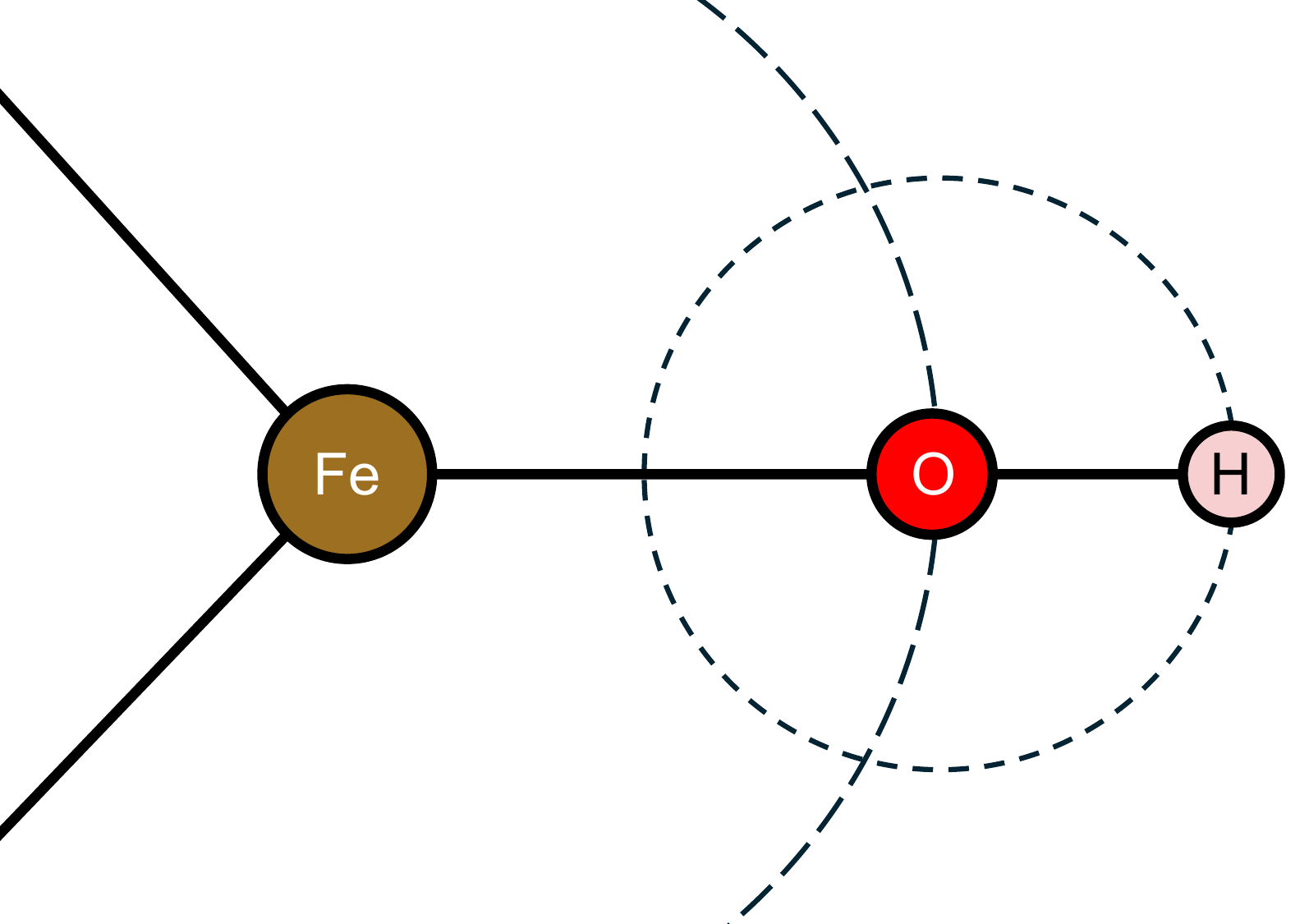}
    \caption{\small\textbf{Asymmetry of Nearest Neighbor Radii.}
        Schematic visualization of a hypothetical iron-hydroxide bonding arrangement.
        Atoms are depicted as circles, nearest neighbor bonds as black lines.
        The nearest neighbor radius of the oxygen atom (small dashed circle) does not encompass iron, while the iron nearest neighbor radius (large dashed circle) encompasses oxygen.
    }
    \label{figS3}
\end{figure*}

\begin{figure*}[ht]
    \centering
    \includegraphics[width=0.55\linewidth]{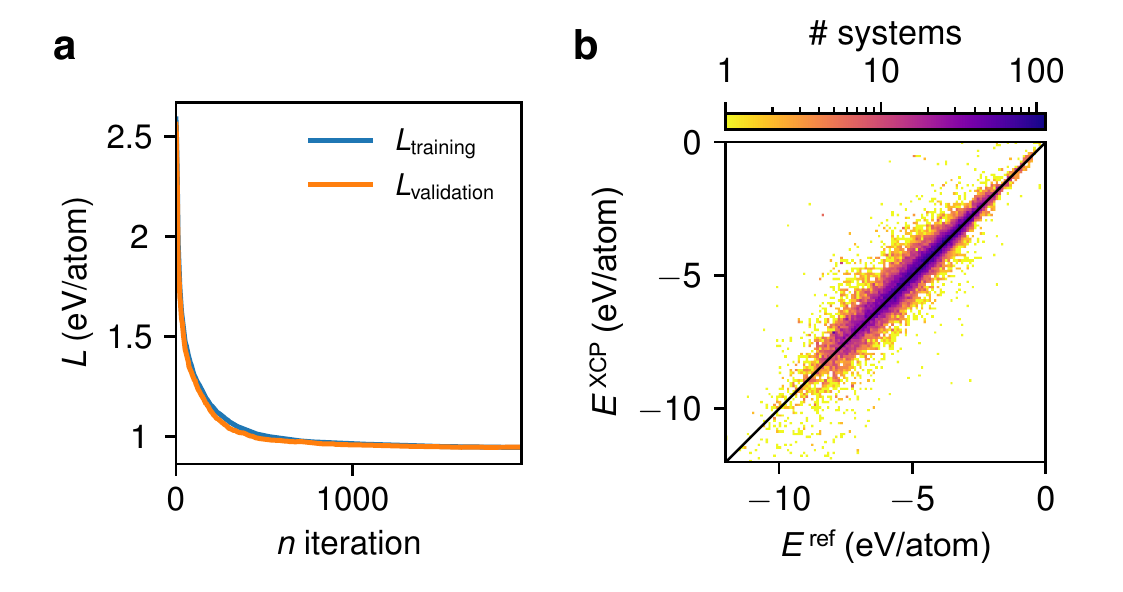}
    \caption{\small\textbf{Loss Minimization.}
        (\textbf{a}) Training and validation loss during potential optimization.
        (\textbf{b}) Reference versus model energies for the final potential.
    }
    \label{figS4}
\end{figure*}

\begin{figure*}[ht]
    \centering
    \includegraphics[width=0.5\linewidth]{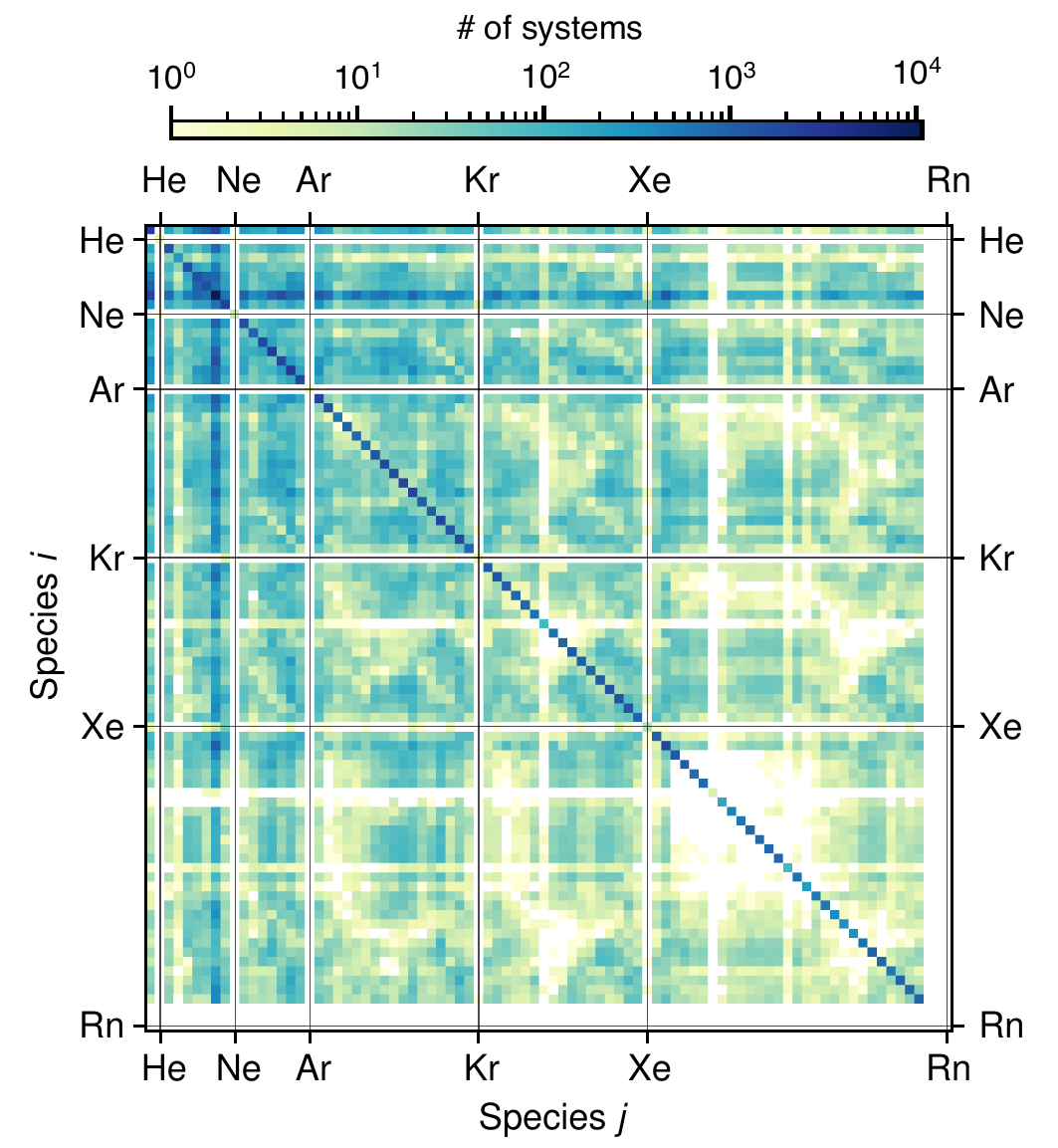}
	\caption{\small\textbf{Representation Bias.}
    	Distribution of bond pairs in the AFLOW-ICSD dataset. 
    	Each square in the matrix corresponds to a species pair $(i,j)$, with the color indicating the number of systems exhibiting this interaction according to the neighbor determination criterion developed in this work. 
    	Systems containing multiple bond types are counted for each pair, \emph{i.e.}, the total count exceeds the number of distinct systems.
	}
    \label{figS5}
\end{figure*}

\begin{figure*}[ht]
    \centering
    \includegraphics[width=0.5\linewidth]{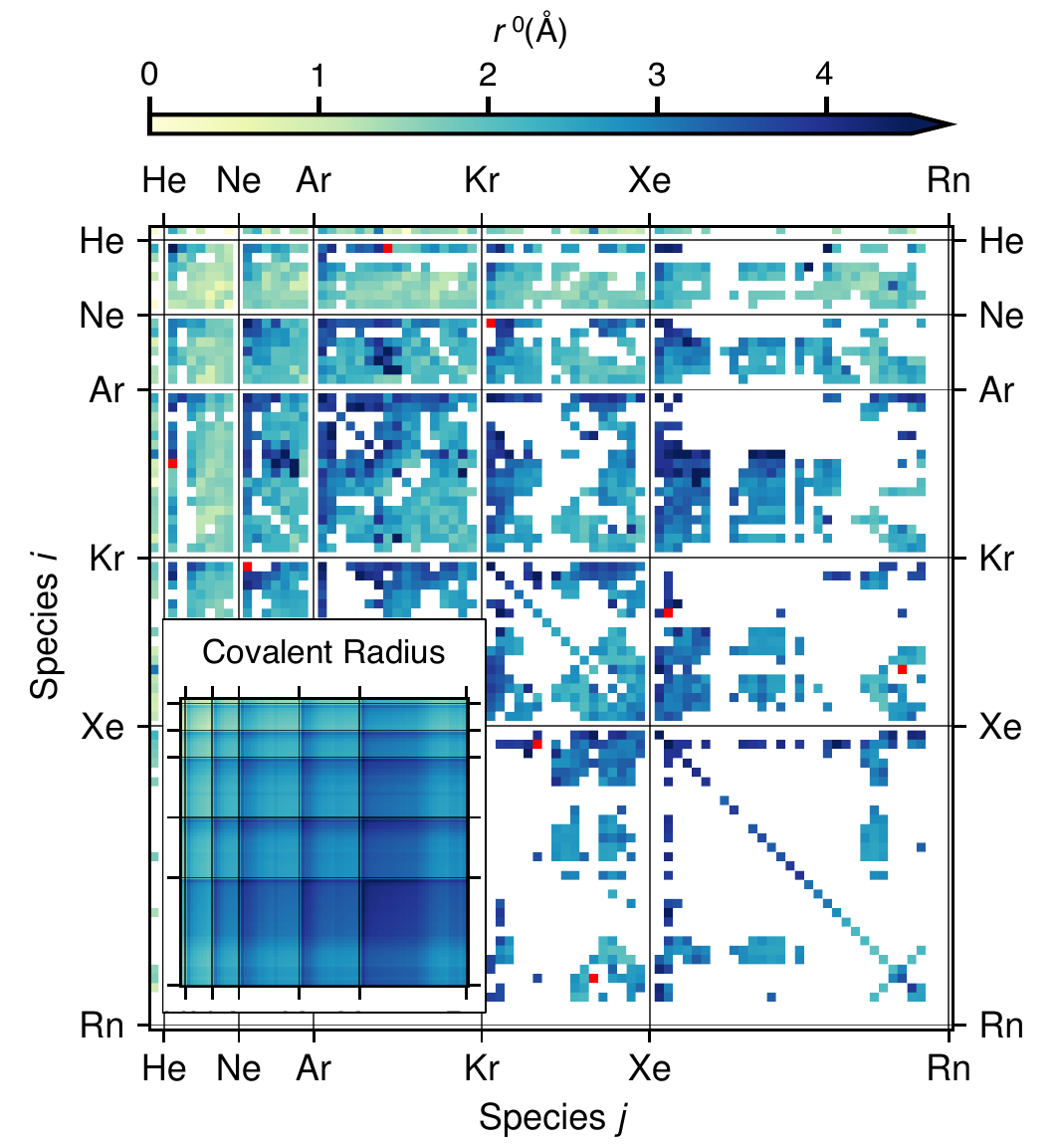}
    \caption{\small\textbf{Equilibrium Radii.}
		Visualization of the numerically determined equilibrium radius of the fitted two-body potential, \emph{i.e.}, $\operatorname{argmin}V(r)$.
		Each square in the matrix corresponds to a species pair $(i,j)$, with the color scheme indicating the equilibrium bonding radii according to the adjacent bar.
		Missing data is shown in white; red squares mark pairs with no equilibrium radius. Data is symmetric about the diagonal as bonding is reciprocal.
		The inset shows the sum of covalent radii from Ref.~\cite{Cordero_DT_2008}.
	}
    \label{figS6}
\end{figure*}

\begin{figure*}[ht]
    \centering
    \includegraphics[width=0.45\linewidth]{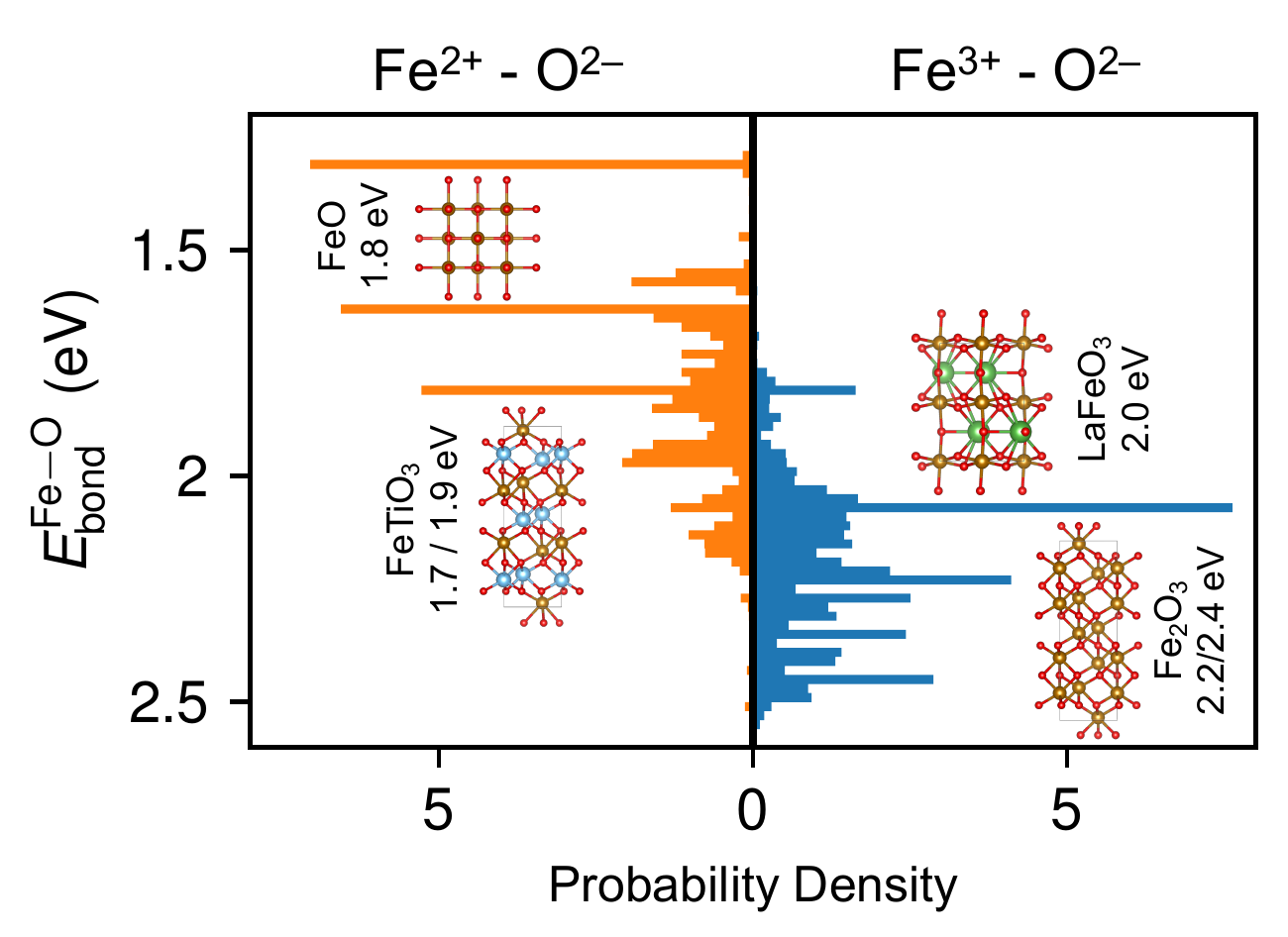}
    \caption{\small\textbf{Different Chemical Environments.}
    	Probability density of Fe--O bonding energies $E_\mathrm{bond}^\mathrm{Fe-O}$ in the AFLOW-ICSD dataset for Fe in the +2 and +3 oxidation states.
    	As an inset a range of representative structures for the two oxidation states together with their Fe--O bonding energies are displayed.
    	If a structure has two different bonding energies, \emph{e.g.}, for short and long bonds, all energies are shown.
    	Oxidation states are determined on a database scale using the approaches developed for the CCE framework~\cite{Friedrich_CCE_2019}. 
    }
    \label{figS7}
\end{figure*}

\begin{figure*}[ht]
    \centering
    \includegraphics[width=.7\linewidth]{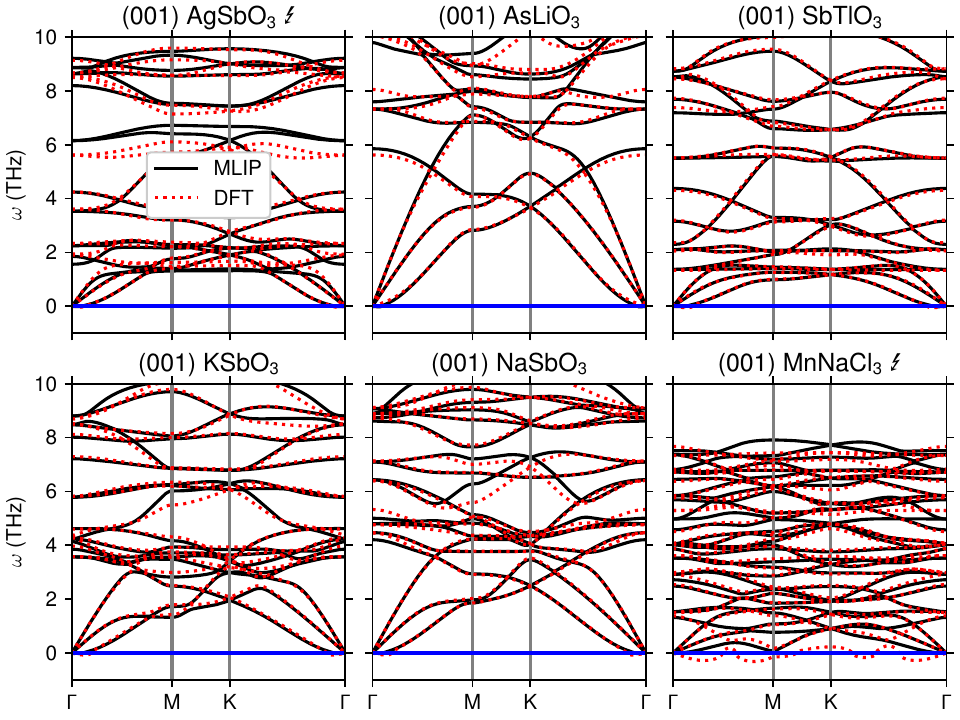}
    \caption{\small\textbf{Phonon Calculation Benchmark.}
      \textit{Ab-initio} phonon dispersions for non-vdW 2D candidates obtained from Ref.~\cite{Barnowsky_AdvElMats_2023} (shown in red) in comparison to phonon dispersions determined using the mace-mh-1 MLIP (shown in black).
      Materials for which there is a large mismatch ($\epsilon > 0.1$) between the DFT and MLIP ground state structures are labeled with a lightning icon.
    }
    \label{figS8}
\end{figure*}

\FloatBarrier

\section{Supplementary Tables}
\vspace{-.3cm}
\begin{table*}[ht]
   \centering
    \caption[]{\small\textbf{Benchmark of the XCP Method with Respect to Experimentally Exfoliated Systems.}
            List of experimentally exfoliated non-vdW 2D materials, limited to studies where both the exfoliated plane and structure are explicitly stated in the original publication. 
            The table reports the bulk compound, Miller indices of the exfoliated lattice plane, exfoliation technique (L = liquid-phase, M = mechanical), literature reference, as well as the bulk structure's ICSD database identifier, space group, and Pearson symbol.
            Entries are sorted by publication date.
            The last two columns indicate whether the exfoliated planes were correctly predicted by the HKLSEARCH (H) and BONDDEL (B) approaches.
            For successful matches, the corresponding exfoliability ratio is listed.
            The final row summarizes the overall success rate if both approaches are combined.
    }
    \label{tabS1}
    \vspace{1em}
    \begin{tabular}{l|c|c|c|c|c|c|c|c}
        \hline
        Material & Plane & Technique & Ref. & Bulk \# ICSD & Bulk Space Group & Bulk Pearson Symbol & $R_{\mathrm{H}}$ & $R_{\mathrm{B}}$ \\\hline\hline
        $\alpha$-WO$_3$            & $(001)$                  & L         & \cite{Guan_AdvMat_2017}             & 27961   & $P4/nmm$ (\#129)            & tP8   & 1.15 & 1.34 \\\hline
        LiMn$_2$O$_4$              & $(111)$                  & L         & \cite{Tai_advmat_2017}              & 53603   & $Fd\overline{3}m$ (\#227)   & cF56  & 1.15 &      \\\hline
        LiFePO$_4$                 & $(010)$                  & L         & \cite{Tai_advmat_2017}              & 260571  & $Pnma$ (\#62)               & oP28  &      &      \\\hline
        $\beta$-B                  & $(104)$                  & L         & \cite{Li_acsnano_2018}              & 240997  & $R\overline{3}m$ (\#166)    & hR105 & 1.09 &      \\\hline
        $\alpha$-Fe$_2$O$_3$       & $(001)$                  & L         & \cite{Puthirath_Balan_NNANO_2018}   & 161292  & $R\overline{3}c$ (\#167)    & hR10  &      & 1.06 \\\hline
        FeTiO$_3$                  & $(001)$                  & L         & \cite{Puthirath_Balan_CoM_2018}     & 247547  & $R\overline{3}$ (\#148)     & hR10  &      & 1.63 \\\hline
        FeCr$_2$O$_4$              & $(111)$                  & L         & \cite{Yadav_AdvMatInt_2018}         & 183962  & $Fd\overline{3}m$ (\#227)   & cF56  & 1.11 &      \\\hline
        MnTe                       & $(110)$                  & L         & \cite{Puthirath_Balan_acsanm_2018}  & 174028  & $P6_3/mmc$ (\#194)          & hP4   & 1.02 &      \\\hline
        Si                         & $(111)$                  & L         & \cite{Wang_JPCL_2020}                & 652255  & $Fd\overline{3}m$ (\#227)   & cF8   & 1.13 &      \\\hline
        Mg                         & $(001)$                  & L         & \cite{Zhang_angchem_2019}           & 642653  & $P6_3/mmc$ (\#194)          & hP2   & 1.02 & 1.02 \\\hline
        $\alpha$-Ti                & $(001)$                  & L         & \cite{Xie_ACSAMI_2019}              & 44872   & $P6_3/mmc$ (\#194)          & hP2   & 1.01 & 1.05 \\\hline
        PbS                        & $(001)$                  & L         & \cite{Liu_JMCA_2019}                & 169770  & $Fm\overline{3}m$ (\#225)   & cF8   & 1.00 &      \\\hline
        CaCO$_3$                   & $(104)$                  & L         & \cite{Liu_JMCA_2019}                & 79673   & $R\overline{3}c$ (\#167)    & hR10  & 1.82 &      \\\hline
        $\alpha$-Ge                & $(111)$                  & L         & \cite{Feng_matter_2020}             & 44841   & $Fd\overline{3}m$ (\#227)   & cF8   & 1.13 &      \\\hline
        Fe$_3$O$_4$                & $(111)$                  & L         & \cite{Puthirath_Small_2020}         & 249047  & $Fd\overline{3}m$ (\#227)   & cF56  & 1.05 &      \\\hline
        B$_4$C                     & $(100)$                  & L, M      & \cite{Guo_Nanoscale_2021}           & 654971  & $R\overline{3}m$ (\#166)    & hR15  &      &      \\\hline
        Sn                         & $(111)$                  & L         & \cite{Ouyang_NML_2021}              & 53789   & $Fd\overline{3}m$ (\#227)   & cF8   & 1.13 &      \\\hline
        TiC                        & $(110)$                  & L         & \cite{Hu_SmallStruct_2021}          & 618935  & $Fm\overline{3}m$ (\#225)   & cF8   &      &      \\\hline
        $\alpha$-B                 & $(101)$                  & L          & \cite{Goktuna_PSCE_2021}           & 26487   & $R\overline{3}m$ (\#166)    & hR12  &      &      \\\hline
        AgCrS$_2$                  & $(001)$                  & L          & \cite{Peng_NChem_2021}             & 42396   & $R3m$ (\#160)               & hR4   & 3.18 & 1.25 \\\hline
        InGaS$_3$                  & $(001)$                  & M          & \cite{Toksumakov_NPJ2DM_2022}      & 62929   & $P6_1$ (\#169)              & hP30  &      &      \\\hline
        Bi                         & $(001)$                  & M          & \cite{Jiang_NatSyn_2023}           & 616519  & $R\overline{3}m$ (\#166)    & hP2   &      & 1.26 \\\hline
        Sb                         & $(001)$                  & M          & \cite{Jiang_NatSyn_2023}           & 651490  & $R\overline{3}m$ (\#166)    & hP2   &      & 1.18 \\\hline
        SnO                        & $(001)$                  & M          & \cite{Jiang_NatSyn_2023}           & 16481   & $P4/nmm$ (\#129)            & tP4   & 2.45 & 5.85 \\\hline
        V$_2$O$_5$                 & $(010)$                  & M          & \cite{Jiang_NatSyn_2023}           & 41030   & $Pmmn$ (\#59)               & oP14  & 3.17 & 4.67 \\\hline
        Bi$_2$O$_2$Se              & $(001)$                  & M          & \cite{Jiang_NatSyn_2023}           & 411143  & $I4/mmm$ (\#139)            & tI10  & 1.45 & 3.42 \\\hline
        CaMn$_4$Si$_5$O$_{15}$     & $(110)$                  & L          & \cite{Mahapatra_2DM_2023}          & 200162  & $P\overline{1}$ (\#2)       & aP50  &      &      \\\hline
        CaF$_2$                    & $(111)$                  & L          & \cite{Chattopadhyay_NL_2024}       & 44937   & $Fm\overline{3}m$ (\#225)   & cF12  & 1.16 &      \\\hline
        Cr$_2$S$_3$                & $(001)$                  & L          & \cite{Su_advsci_2024}              & 16720   & $P\overline{3}1c$ (\#163)   & hP20  & 1.29 &      \\\hline\hline
        \multicolumn{7}{|c|}{$\sum=$}  & \multicolumn{2}{c|}{23/29}\\\hline
    \end{tabular}
\end{table*}

\begin{table*}[ht]
	\caption{\textbf{Pseudopotential Choice.}
		Employed VASP PPs for the PBE functional according to the AFLOW standard.
		Each element is associated with a specific PP which is labeled in accordance with the VASP convention. 
	}
	\label{tabS2}
	\vspace{1em}
    \begin{tabular}{c|c||c|c||c|c||c|c||c|c|}
    	\hline
    	Element & Label & Element & Label & Element & Label & Element & Label & Element & Label\\\hline\hline
Ac & Ac & Cs & Cs\_sv & K & K\_sv & Pa & Pa & Sr & Sr\_sv \\\hline
Ag & Ag & Cu & Cu\_pv & Kr & Kr & Pb & Pb\_d & Ta & Ta\_pv \\\hline
Al & Al & Dy & Dy\_3 & La & La & Pd & Pd\_pv & Tb & Tb\_3 \\\hline
Ar & Ar & Er & Er\_3 & Li & Li\_sv & Pm & Pm & Tc & Tc\_pv \\\hline
As & As & Eu & Eu & Lu & Lu & Pt & Pt & Te & Te \\\hline
Au & Au & F & F & Mg & Mg\_pv & Pr & Pr & Th & Th\_s \\\hline
B & B\_h & Fe & Fe\_pv & Mn & Mn\_pv & Pu & Pu\_s & Ti & Ti\_sv \\\hline
Ba & Ba\_sv & Ga & Ga\_h & Mo & Mo\_pv & Rb & Rb\_sv & Tl & Tl\_d \\\hline
Be & Be\_sv & Ge & Ge\_h & Na & Na\_sv & Re & Re\_pv & Tm & Tm \\\hline
Bi & Bi\_d & Gd & Gd & N & N & Rh & Rh\_pv & U & U \\\hline
Br & Br & H & H & Nb & Nb\_sv & Ru & Ru\_pv & V & V\_sv \\\hline
C & C & He & He & Nd & Nd & S & S & W & W\_sv \\\hline
Ca & Ca\_sv & Hf & Hf\_pv & Ne & Ne & Sb & Sb & Xe & Xe \\\hline
Cd & Cd & Hg & Hg & Ni & Ni\_pv & Sc & Sc\_sv & Y & Y\_sv \\\hline
Ce & Ce & Ho & Ho\_3 & Np & Np\_s & Se & Se & Yb & Yb \\\hline
Cl & Cl & I & I & O & O & Si & Si & Zn & Zn \\\hline
Co & Co & In & In\_d & Os & Os\_pv & Sm & Sm & Zr & Zr\_sv \\\hline
Cr & Cr\_pv & Ir & Ir & P & P & Sn & Sn \\\hline
    \end{tabular}
\end{table*}

\begin{table*}[ht]
	\caption{\textbf{$d$-Element On-Site Parameters.}
		Effective Hubbard $U$ parameters applied to $d$-shell elements.
	}
	\label{tabS3}
	\vspace{1em}
    \begin{tabular}{c|c||c|c||c|c||c|c||c|c|}
    \hline
    Element & $U_\mathrm{eff}$ & Element & $U_\mathrm{eff}$  & Element & $U_\mathrm{eff}$  & Element & $U_\mathrm{eff}$  & Element & $U_\mathrm{eff}$\\\hline\hline
Sc & 2.9 & Ni & 5.1 & Tc & 2.7 & Ta & 2.0 & Hg & 5.1 \\\hline
Ti & 4.4 & Cu & 4.0 & Ru & 3.0 & W & 2.2 & Ga & 3.9 \\\hline
V & 2.7 & Zn & 7.5 & Rh & 3.3 & Re & 2.4 & In & 1.9 \\\hline
Cr & 3.5 & Y & 5.1 & Pd & 3.6 & Os & 2.6 & Bi & 0.0 \\\hline
Mn & 4.0 & Zr & 5.1 & Ag & 4.0 & Ir & 2.8 & Pb & 0.0 \\\hline
Fe & 4.6 & Nb & 2.1 & Cd & 2.1 & Pt & 3.0 & Sn & 3.5 \\\hline
Co & 5.0 & Mo & 2.4 & Hf & 5.1 & Au & 4.0 \\\hline
    \end{tabular}
\end{table*}

\begin{table*}[ht]
	\caption{\textbf{$f$-Element On-Site Parameters.}
		$U$ and $J$ parameters applied to selected $f$-shell elements.
	}
	\label{tabS4}
	\vspace{1em}
    \begin{tabular}{c|c|c||c|c|c}
    \hline
	Element & U & J & Element & U & J\\\hline\hline
La & 8.1 & 0.6 & Tm & 7.0 & 1.0 \\\hline
Ce & 7.0 & 0.7 & Dy & 5.6 & 0.0 \\\hline
Pr & 6.5 & 1.0 & Yb & 7.0 & 0.7 \\\hline
Gd & 6.7 & 0.7 & Lu & 4.8 & 0.9 \\\hline
Nd & 7.2 & 1.0 & U & 4.5 & 0.5 \\\hline
Sm & 7.4 & 1.0 & Th & 5.0 & 0.0 \\\hline
Eu & 6.4 & 1.0 \\\hline
    \end{tabular}
\end{table*}

\FloatBarrier
\end{document}